\documentclass[sigconf]{acmart}
\AtBeginDocument{%
  \providecommand\BibTeX{{%
    \normalfont B\kern-0.5em{\scshape i\kern-0.25em b}\kern-0.8em\TeX}}}
\setcopyright{acmcopyright}
\copyrightyear{2018}
\acmYear{2018}
\acmDOI{XXXXXXX.XXXXXXX}
\acmConference[ASE 2024]{IEEE/ACM International Conference on Automated Software Engineering}{Sun 27 October - Fri 1 November, 2024}{Sacramento, US}
\acmPrice{15.00}
\acmISBN{978-1-4503-XXXX-X/18/06}
\usepackage{tabularx}
\usepackage{color}
\usepackage{multirow}
\usepackage{graphicx}
\usepackage{subcaption}
\usepackage{array}
\usepackage{tcolorbox}
\usepackage{xspace}
\usepackage{colortbl}
\usepackage{xspace}
\usepackage{mdframed}
\usepackage{enumitem}
\usepackage{appendix}
\definecolor{background}{HTML}{EBFEED}
\definecolor{edge}{HTML}{B2DDA3}
\definecolor{morandiRed}{RGB}{255, 220, 220}
\definecolor{morandiGreen}{RGB}{102, 153, 136}
\newtcolorbox{mybox}{
colback=background!50, colframe=edge,
width=\columnwidth,
arc=1.2mm,
auto
outer
arc
}

\newcommand\llama{\textsc{Llama}}
\newcommand\llamafull{\textsc{Llama\Spacing{}\Hyphen{}2\Spacing{}\Hyphen{}7B}}
\newcommand\codellamaTerm{Code\Spacing{}Llama}
\newcommand\codellama{\textsc{\codellamaTerm}}
\newcommand\codellamasmall{\textsc{\codellamaTerm\Spacing{}\Hyphen{}7B}}
\newcommand\codellamalarge{\textsc{\codellamaTerm\Spacing{}\Hyphen{}13B}}
\newcommand\Spacing{}
\newcommand\Hyphen{\Spacing{}-\Spacing{}}
\newcommand\deepseekterm{Deep\Spacing{}Seek}
\newcommand\deepseekcoder{\deepseekterm\Spacing{}\Hyphen{}Coder}
\newcommand\deepseekfull{\deepseekterm\Spacing{}\Hyphen{}Coder\Spacing{}\Hyphen{}Base\Spacing{}\Hyphen{}6.7B}
\newcommand\aprinst{\textsc{APR\Spacing{}\Hyphen{}Instruction}}
\newcommand\ossinst{\textsc{OSS\Spacing{}\Hyphen{}Instruction}}
\newcommand\ossinstpart{\textsc{OSS\Spacing{}\Hyphen{}Instruction}\Spacing{}\Hyphen{}30K}
\newcounter{finding}
\newcommand{\finding}[1]{\refstepcounter{finding}
 	\vspace{3mm}
	\begin{mdframed}[linecolor=gray,roundcorner=12pt,backgroundcolor=gray!15,linewidth=3pt,innerleftmargin=2pt, leftmargin=0cm,rightmargin=0cm,topline=false,bottomline=false,rightline=false]
	\textbf{Finding \arabic{finding}:} #1
	\end{mdframed}
}
\begin{document}

\title{Exploring Parameter-Efficient Fine-Tuning of Large Language Model on Automated Program Repair}

\author{Guochang Li}
\affiliation{%
  \institution{Zhejiang University}
  \city{Hangzhou}
  \country{China}}
\email{gcli@zju.edu.cn}

\author{Chen Zhi}
\authornote{Corresponding authors.}
\affiliation{ 
  \institution{Zhejiang University}
  \city{Hangzhou}
  \country{China}}
\email{zjuzhichen@zju.edu.cn}

\author{Jialiang Chen}
\affiliation{%
  \institution{Zhejiang University}
  \city{Hangzhou}
  \country{China}}
\email{cjl99@zju.edu.cn}

\author{Junxiao Han}
\affiliation{%
  \institution{Hangzhou City University}
  \city{Hangzhou}
  \country{China}}
\email{hanjx@hzcu.edu.cn}

\author{Shuiguang Deng}
\authornotemark[1]
\affiliation{%
  \institution{Zhejiang University}
  \city{Hangzhou}
  \country{China}}
\email{dengsg@zju.edu.cn}

\renewcommand{\shortauthors}{Li and Zhi, et al.}

\begin{abstract}

Automated Program Repair (APR) aims to fix bugs by generating patches. And existing work has demonstrated that ``pre-training and fine-tuning'' paradigm enables Large Language Models (LLMs) improve fixing capabilities on APR. However,  existing work mainly focuses on Full-Model Fine-Tuning (FMFT) for APR and limited research has been conducted on the execution-based evaluation of Parameter-Efficient Fine-Tuning (PEFT) for APR. Comparing to FMFT, PEFT can reduce computing resource consumption without compromising performance and has been widely adopted to other software engineering tasks.

To fill this gap, we enhance the existing APR dataset by employing prompt engineering to create an instruction dataset, \aprinst{}, at first. Secondly, we fine-tune four pre-trained LLMs using four different PEFT methods with \aprinst{}. The best fine-tuned model fixes 58\% more bugs than the state-of-the-art LLM-based APR techniques. The results also show that $(IA)^3$ improves the creativity of LLMs more effectively through fine-tuning and achieves the highest fixing capability compared to the other three PEFT methods. Thirdly, we explore the optimal configuration of PEFT hyperparameters, and assess the impact of instruction dataset size, showing that a larger number of parameters and a larger training dataset do not necessarily result in better performance for PEFT. Lastly, we analyze peak memory usage and trainable parameters to show the efficiency of PEFT.

This work provides a comprehensive exploration of PEFT on APR and suggests potentially promising directions for extension to other software engineering downstream tasks. \aprinst{}, PEFT weights, and the fine-tuning code are publicly available as open-source resources.
\end{abstract}


\maketitle

\section{Introduction}
\label{sec:intro}

Automated Program Repair (APR) aims to generate patches for program bugs automatically, without human intervention. Based on existing studies \cite{benton2020effectiveness, liu2020efficiency, wang2020automated, zhang2023survey, huang2023survey}, traditional APR techniques include search-based \cite{le2011genprog, jiang2018shaping}, constraint-based \cite{nguyen2013semfix, xuan2016nopol} and template-based \cite{le2016history, liu2019you}, which require experts to provide prior knowledge manually. Learning-based APR techniques \cite{chen2019sequencer, lutellier2020coconut}, such as Neural Machine Translation (NMT) techniques \cite{tufano2019empirical}, enhance fixing capabilities through translating bug lines into fix lines. Nowadays, LLM-based APR techniques \cite{xia2022less} have gained significant attention and are regarded as highly promising APR techniques.

Some existing studies have employed LLMs directly on APR. For instance, Alpharepair \cite{xia2022less} enhances fixing capabilities by directly predicting the fix line based on the provided bug line context. While zero-shot can already aids in completing certain APR tasks, Full-Model Fine-Tuning (FMFT) LLMs with domain-specific datasets can fully unlock its potential \cite{radford2019language, wei2021finetuned, huang2023empirical}. Recently, researchers have undertaken FMFT with APR datasets, which shows significant improvements on fixing capabilities of LLMs, enabling them to fix 46\% to 164\% more bugs compared to existing deep learning (DL)-based APR techniques \cite{jiang2023impact}. 

Instruction-tuning \cite{wang2023self} enables LLMs to align the training objectives with downstream tasks. However, instruction datasets \cite{codealpaca, wei2023magicoder} of software engineering (SE) mainly consist of general code generation tasks, such nature language to program language. There is a lack of APR-specific instruction datasets for APR tasks. Although some APR datasets are available for DL-based APR techniques \cite{zhu2021syntax}, which cannot be used for instruction-tuning due to the limited variety of instruction types. To adapt to instruction-tuning, it is necessary to enrich the instruction types, such as descriptive information, of existing datasets to build an APR instruction dataset.

FMFT demands significant computing resources which may be not practical for all users \cite{liu2024delving}. In contrast, PEFT methods \cite{hu2021lora, li2021prefix, liu2021p, liu2022few} have emerged as a more promising approach for limited computing resources \cite{weyssow2023exploring}. PEFT guarantees the preservation of the core generation capability of LLMs by freezing the initial parameters and fine-tuning additional parameters, which do not exceed 1\%-5\% of the initial parameters \cite{liu2024delving}. Previous studies \cite{choi2023codeprompt, zou2023comprehensive} have demonstrated the superiority of PEFT in specific software engineering (SE) downstream tasks. However, there is currently no research investigating PEFT methods on APR. 

Several SE tasks that employ PEFT, such as code completion \cite{chen2021evaluating}, docstring generation \cite{lu2021codexglue}, and code clone detection \cite{zou2023comprehensive}, are evaluated by Natural Language Processing (NLP) metrics like BLEU \cite{papineni2002bleu}. These metrics do not reliably measure the functional correctness of the generated code \cite{xia2023automated, liu2024your}. Unlike other SE tasks, APR requires generated patches that can be merged into the original code and executed correctly. Therefore, while generated code segments may exhibit high similarity, it does not ensure that executing the two code segments will produce similar or correct results. Evaluation of the generated patches needs to be performed using execution-based methods, such as test cases. Limited research has focused on the execution-based evaluation of PEFT results, although execution-based evaluation plays a crucial role in numerous code generation tasks involving APR.

To address the above issues, this work comprehensively explores PEFT on APR, with a focus on the construction of \aprinst, supervised fine-tuning (SFT) with PEFT, and patch validation. To begin with, we combine prompt engineering with existing APR datasets \cite{zhu2021syntax} to enrich additional descriptive information, such as problem descriptions and bug causes, resulting \aprinst, which consists of 30,000 instructions and has been made publicly available \cite{repo}. In order to assess the effectiveness, a comparative analysis is conducted with two widely used SE instruction datasets, Code Alpaca \cite{codealpaca} and \ossinst{} \cite{wei2023magicoder}. The results of experiments provide evidence that \aprinst{} \cite{repo} holds better potential for improving fixing capabilities of LLMs, which fixes 20.1\% more than the other two datasets at most.

Secondly, we conduct SFT using \textit{LoRA} \cite{hu2021lora}, \textit{prefix-tuning} \cite{li2021prefix}, \textit{p-tuing v2} \cite{liu2021p}, $(IA)^3$ \cite{liu2022few} on \codellamasmall{} \cite{codellama}, \codellamalarge{} \cite{codellama}, \llamafull{} \cite{llama2}, \deepseekfull{} \cite{deepseek-coder} with \aprinst{}, and evaluating on Defects4J v2.0 \cite{just2014defects4j}, QuixBugs \cite{lin2017quixbugs}, and HumanEval-Java \cite{jiang2023impact} to showcase the effectiveness and efficiency of PEFT in the single-hunk APR task. Initially, we compare fixing capabilities of PEFT with the state-of-the-art (SOTA) APR techniques, which demonstrates that PEFT can at most fix 58\% more bugs than the SOTA APR techniques. Subsequently, we investigate fixing capabilities of no fine-tuning as the baseline, \textit{LoRA} \cite{hu2021lora}, \textit{prefix-tuning} \cite{li2021prefix}, \textit{p-tuing v2} \cite{liu2021p}, $(IA)^3$ \cite{liu2022few}, and FMFT on \codellamasmall. The results suggest that, compared to no fine-tuning, PEFT methods enhance the fixing capabilities of LLMs, fixing 7.2\%-31.3\% more bugs. 

Among four LLMs and four PEFT methods in this work, \deepseekfull{} achieve the highest fixing capability following $(IA)^3$. Furthermore, we try to analyze why $(IA)^3$ achieve the best performance among four PEFT methods. Combining $pass@k$ and an instance, we demonstrate that $(IA)^3$ improves the creativity of LLMs more effectively through fine-tuning compared to the other three PEFT methods, which accounts for its superior performance. PEFT methods alsoreduce around 45.6\%-50.0\% peak GPU memory usage compared to FMFT, which shows the efficiency of PEFT. 

\begin{figure*}[t]
  \centering
  \includegraphics[width=\textwidth]{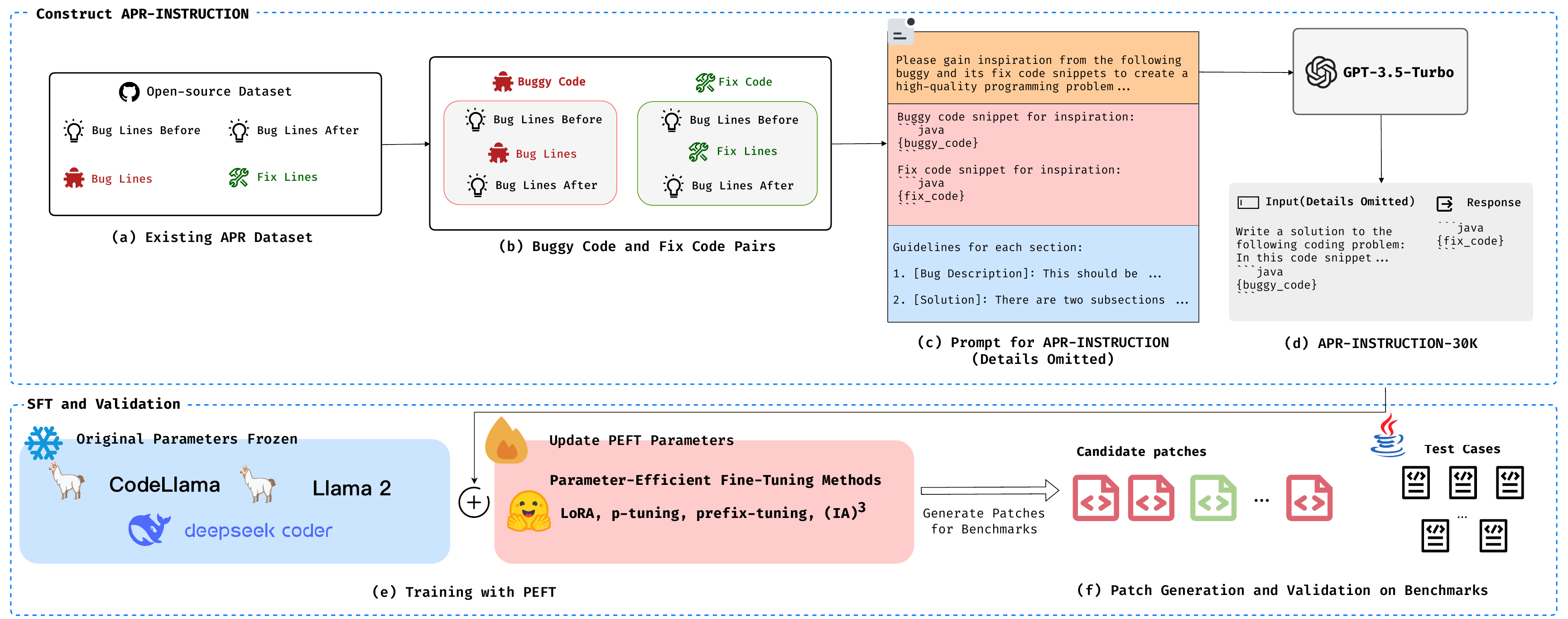}
  \caption{An overview of \aprinst{} constructed, supervised fine-tuning on four LLMs with four PEFT Methods, and Evaluation on three benchmarks}
  \label{fig:arch}
\end{figure*}

Finally, we conduct further investigations into two aspects aimed at improving the efficiency of PEFT: hyperparameters and the size of instruction datasets. We examine the impact of hyperparameters, such as \textit{rank} of \textit{LoRA} on APR, since a lower \textit{rank} implies a reduction of trainable parameters, resulting in less memory usage. The results indicate that setting \textit{rank} of \textit{LoRA} to $16$ already enables LLMs to achieve a relatively strong fixing capability. Increasing the \textit{rank} blindly does not necessarily yield further improvements in fixing capabilities. We examine the impact of instruction dataset size, considering that a smaller dataset implies reduced computational resource costs. To investigate this, we employ $(IA)^3$ to fine-tune different sizes of \aprinst{} dataset, revealing $(IA)^3$ only fixes 3.67\% less bugs when reducing 50\% instructions. This suggests the feasibility of appropriately reducing the instruction dataset when resources are limited.

To sum up, the main contributions of this work are as follows:

\begin{itemize}[topsep=0pt]
    \item This work constructs \aprinst{} by enhancing the existing APR training dataset by adding extra descriptions, which indeed improves fixing capabilities of LLMs on APR.
    \item This work shows \codellamasmall{} with all PEFT methods fixes 7.2\%-31.3\% more bugs than no fine-tuning, and reduce around around 45.6\%-50.0\% peak GPU memory usage compared to FMFT, showing the effectiveness and efficiency of PEFT on APR.
    \item This work shows \deepseekfull{} with $(IA)^3$ achieves the best fixing capability, fixing 58\% more bugs than the SOTA technique. 
    \item This work investigates multiple factors that potentially influence the fixing capability of PEFT methods, including scales of LLMs, type of pre-training data, and base models.
    \item This work demonstrates that $(IA)^3$ achieve the best fixing capability, benifiting from its superior creativity improvement through SFT compared to the other three PEFT methods, possibly.
    \item This work further explores the impact of hyperparameters of PEFT and the size of instruction dataset to improve the efficiency of PEFT.
    \item This work, including \aprinst{}, code, and weights, has been publicly released on \cite{repo} to allow researchers to further extend it to other software engineering tasks.
\end{itemize}

Overall, this work primarily focuses on investigating the effectiveness and efficiency of PEFT on APR, which presents a comprehensive roadmap, offering complete and reproducible data. By applying the same workflow, PEFT methods have the potential to be extended to other SE tasks by leveraging existing datasets in order to enhance performance of downstream tasks with LLMs.

\section{Experiment Design}
\label{sec:experiment_design}

\subsection{Overview}
\label{subsec:overview}

Figure \ref{fig:arch} illustrates the overview of this work, which consists of three main stages: constructing \aprinst, supervised fine-tuning (SFT) with PEFT, and generating and validating patches. 

In the initial stage of constructing \aprinst, we utilize an APR dataset previously released \cite{zhu2021syntax} as Figure \ref{fig:arch} (a) shown. This dataset contains pairs of bug code and fix code as Figure \ref{fig:arch} (b) shown, served as seed fragments. To enhance the diversity of \aprinst, we employ prompt engineering with ChatGPT (i.e., the GPT-3.5-turbo model with its default setting \cite{chatgpt}) as Figure \ref{fig:arch} (c) shown to incorporate additional descriptive information as Figure \ref{fig:arch} (d) shown. In the second stage, we perform SFT on four LLMs with four PEFT methods based on \aprinst{} as Figure \ref{fig:arch} (e) shown. The training weights are saved for inference on benchmarks, and for each bug code, we generate 10 candidate patches. Finally, we assess the correctness of the generated patches by test cases in benchmarks as Figure \ref{fig:arch} (f) shown.

\subsection{Studied Large Language Models}

We select LLMs for this work on the following criteria: (1) since the currently available open-source LLMs are predominantly built with a decoder-only architecture \cite{zhang2024systematic}, we specifically focus on selecting open-source LLMs that adhere to this structure. (2) LLMs such as \llama{} 2 \cite{llama2} typically have at least 7 billion (7B) parameters. Hence, all LLMs in this work have sizes around 7B or larger. (3) LLMs are obtained from \textit{Hugging Face} \cite{huggingface}, which are publicly accessible. Finally, we select four LLMs: \codellamasmall{} \cite{codellama}, \codellamalarge{} \cite{codellama}, \llamafull{} \cite{llama2} and \deepseekfull{} \cite{deepseek-coder}

\llama{} 2 \cite{llama2} is pre-trained on the next token prediction task and incorporates several architectural advancements compared to \llama{} \cite{touvron2023llama}, including an increased context length from 2K to 4K. Pre-training data for \llama{} 2 comprises a new mixture of publicly available online sources, resulting in a corpus size of 2 trillion tokens. In this work, we focus on \llamafull{}. \codellama{} \cite{codellama} is a family of LLMs designed for code generation and infilling. It is achieved by further training \llama{} 2 on code-specific datasets, using a dataset that comprises 500 billion tokens. \codellama{} is specifically trained on the fill-in-the-middle task, aimed at generating code that closely matches an existing prefix and suffix. The context window of \codellama{} is 16K. In this work, we focus on analyzing \codellamasmall{} and \codellamalarge{}. \deepseekcoder{} \cite{deepseek-coder} undergoes pre-training with two main objectives: next token prediction and fill-in-the-middle. The training dataset for \deepseekcoder{} consists of 87\% source code, encompassing 603 million files from 87 programming languages and comprising 2 trillion tokens. In this work, we concentrate on \deepseekfull{}.

\subsection{\aprinst{} Construction}

\subsubsection{Data Source}

\aprinst{} is derived from the APR dataset shared in the previous research \cite{zhu2021syntax}, which is obtained from commits of open-source Java projects on GitHub. Each individual fix in the dataset is treated as a separate instance, resulting in a total of 143,666 instances \cite{zhu2021syntax}. Previous studies on APR have predominantly utilized FMFT, and this dataset has been employed to create pairs of bug code and fix code for training \cite{zhu2021syntax,jiang2023impact}. The format of the dataset is illustrated in Figure \ref{fig:arch} (a). However, since instruction-tuning typically requires a diverse range of instruction types to achieve optimal results, this dataset cannot be directly used for instruction-tuning. Additional methods are necessary to construct a more varied instruction dataset.

\subsubsection{Data Construction}

Considering the cost, we employed \textit{gpt-3.5-turbo} \cite{chatgpt} for generating \aprinst. To avoid the attention of LLMs dispersed by long context, we firstly tokenize buggy codes by \codellama{} tokenizer. We only retain buggy codes with tokenized vector lengths not exceed 200. By utilizing the prompt template shown in Figure \ref{fig:arch} (c), we employ \textit{gpt-3.5-turbo} to generate extra descriptions of the original code, including problem description, bug description and so on. The overall process results in a dataset, \aprinst{} containing 30,000 APR instructions \cite{repo}.

\subsubsection{Impact of \aprinst}

To demonstrate the effectiveness of \aprinst{}, a comparison is conducted on two datasets, Code Alpaca \cite{codealpaca} and \ossinst{} \cite{wei2023magicoder}, which are widely used in instruction-tuning for software engineering tasks. SFT is performed on two base models, \codellamasmall{} and \deepseekfull{}, to eliminate the influence of the base model on fixing capabilities. Additionally, in order to exclude the impact of the PEFT method on fixing capabilities, two PEFT methods, \textit{LoRA} and \textit{p-tuning}, are applied to \deepseekfull{}. \ossinst{} consists of 75,000 instructions. To mitigate the impact of data volume for SFT, a random selection of 30,000 instructions from \ossinst{} was executed to construct the \ossinstpart{} for instruction-tuning. On the other hand, Code Alpaca comprises a total of 20,000 instructions, which were totally used for instruction-tuning. After SFT, LLMs generate patches on three benchmarks, which are assessed correctness by test cases. Results are presented in Table \ref{tab:dataset_results}.

\begin{table}[H]
  \caption{Number of fixed bugs by three instruction datasets, two LLMs and two PEFT methods. The best results are highlighted in red.}
  \label{tab:dataset_results}
  \vspace{-3mm}
  \resizebox{0.48\textwidth}{!}{%
  \begin{tabular}{cccccc}
  \hline
  \multirow{3}{*}{\begin{tabular}[c]{@{}c@{}}Base Model\\ with \\ PEFT Method \end{tabular}} & \multirow{3}{*}{Instruction Dataset} & \multicolumn{3}{c}{Benchmark} & \multirow{3}{*}{\begin{tabular}[c]{@{}c@{}}Total\\ Fixed \\ Bugs\end{tabular}} \\ \cline{3-5} 
  & & \begin{tabular}[c]{@{}c@{}}HumanEval-Java\\ (pass@10)\end{tabular} & \begin{tabular}[c]{@{}c@{}}Defects4J v2.0\\ (pass@10)\end{tabular} & \begin{tabular}[c]{@{}c@{}}QuixBugs\\ (pass@10)\end{tabular} \\ \hline
  \multirow{3}{*}{\begin{tabular}[c]{@{}c@{}}\codellamasmall\\ \textit{LoRA} \end{tabular}} & \ossinstpart & 97/163 & 46/217 & 28/40 & 171/420 \\
  & Code Alpaca & 95/163 & 56/217 & 33/40 & 184/420 \\
  & \cellcolor{morandiRed}{\aprinst} & 98/163 & 75/217 & 26/40 & \cellcolor{morandiRed}{199/420} \\ \hline
  \multirow{3}{*}{\begin{tabular}[c]{@{}c@{}}\deepseekcoder \\ Base 6.7B \\ \textit{LoRA}\end{tabular}} & \ossinstpart & 109/163 & 45/217 & 34/40 & 188/420 \\
  & Code Alpaca & 101/163 & 59/217 & 32/40 & 192/420 \\
  & \cellcolor{morandiRed}{\aprinst} & 109/163 & 92/217 & 30/40 & \cellcolor{morandiRed}{231/420} \\ \hline
  \multirow{3}{*}{\begin{tabular}[c]{@{}c@{}}\deepseekcoder \\ Base 6.7B \\ \textit{p-tuning}\end{tabular}} & \ossinstpart & 102/163 & 50/217 & 34/40 & 186/420 \\
  & Code Alpaca & 96/163 & 56/217 & 30/40 & 182/420 \\
  & \cellcolor{morandiRed}{\aprinst} & 108/163 & 84/217 & 33/40 & \cellcolor{morandiRed}{225/420} \\ \hline
  \end{tabular}
  }
  \end{table}

Table \ref{tab:dataset_results} demonstrates that \aprinst{} outperforms \ossinst{} and Code Alpaca in terms of fixed bugs across two LLMs and two PEFT methods, indicating its improved fixing capabilities with instruction-tuning. Specifically, \codellamasmall{} with LoRA fixed an additional 28 bugs with \aprinst{} than \ossinst{} with the lowest performance, and addressed 15 more bugs than Code Alpaca. Similarly, after fine-tuning, \deepseekfull{} using LoRA with \aprinst{}  achieves the highest fixing capability, which eliminate the influence of base models on the improvement of fixing capabilities. Furthermore, based on \deepseekfull{}, instruction-tuning with p-tuning, along with \aprinst{}, demonstrates the highest fixing capability, which eliminates the influence of PEFT methods. In conclusion, \aprinst{} significantly enhances fixing capabilities through fine-tuning. To promote wider adoption by researchers, we have open-sourced \aprinst{} \cite{repo}.

\subsubsection{Evaluation of overlap between \aprinst{} and benchmarks}

In conclusion, there is no overlapping between the instruction dataset \aprinst{} and three benchmarks, namely HumanEval-Java, Quixbugs, and Defects4J. On the one hand, the collection of \aprinst{}  has excluded these benchmarks. \aprinst{} is constructed on an existing dataset \cite{zhu2021syntax}, which claims the dataset is collected from Java projects created on GitHub between March 2011 and March 2018 and excludes projects that are clones of Defects4J projects or use Defects4J, ensuring no overlap with the Defects4J. Humaneval-Java project, with its first commit in 2023, is not included in the collection period of the existing dataset \cite{zhu2021syntax}, ensuring no overlap with the existing APR dataset \cite{zhu2021syntax}. \aprinst{} uses GPT to add some additional description to this existing APR dataset \cite{zhu2021syntax} without more code added. On the other hand, prior work \cite{zhu2021syntax,jiang2023impact} uses this existing APR dataset \cite{zhu2021syntax} for training and also conducts evaluation on Defects4J, QuixBugs, and Humaneval-Java, further indicating no data overlap between \aprinst{} and these three benchmarks.
 
\subsubsection{Evaluation of \aprinst{} validity}

\aprinst{} contains some information generated by GPT-3.5, so we conduct the evaluation for the validity of \aprinst. In the process of generating data, we initially produce nearly 100 samples and conduct a manual inspection, finding no anomalies. Subsequently, we generate the final set of 30k instructions. Given the large size of \aprinst, it is impractical to verify each instruction individually. Even if there is some inaccurate data, prior studies \cite{wei2023magicoder, honovich2023unnatural} suggest that not removing noisy data can actually improve performance. The study \cite{maharana2024mathbb} posits that low-quality data can still positively contribute to the diversity of the dataset. Therefore, we do not necessarily need to remove it. Overall, the impact of data validity is manageable and does not significantly affect the results.

\subsection{PEFT Methods and Implementation Details}

\begin{figure}[H]
  \centering
  \includegraphics[width=0.35\textwidth]{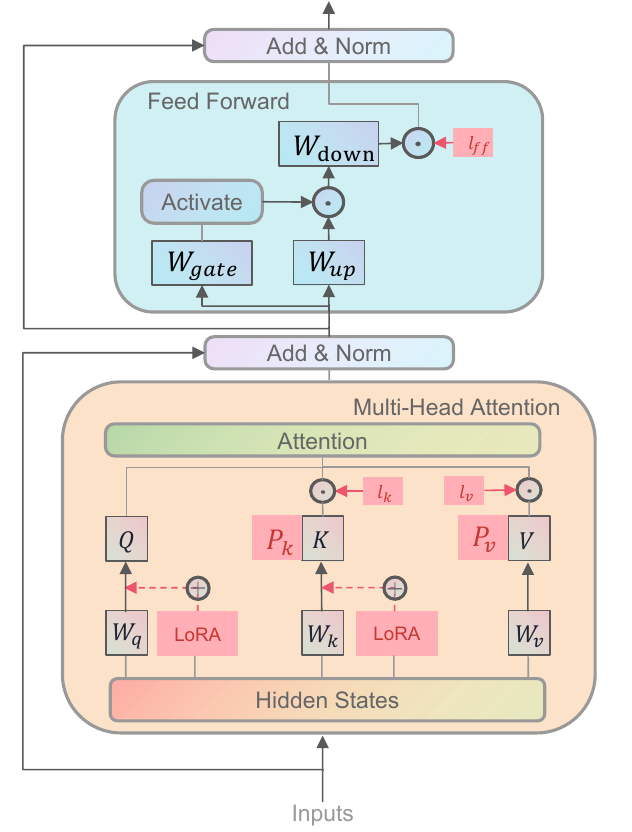}
  \caption{Principles of LoRA, p-tuning, prefix-tuning, and $(IA)^3$ on LLMs. Red parts represents trainable modules of PEFT methods.}
  \label{fig:peft_arch}
\end{figure}

PEFT library \cite{peft} developed by \textit{Hugging Face} team, encompasses multiple widely used PEFT methods. Additionally, open-source LLMs released on \textit{Hugging Face}, can also be easily used through Transformers library \cite{wolf-etal-2020-transformers}. Therefore, combining PEFT library with Transformers library provides a convenient approach for importing and then fine-tuning LLMs. As a result, this work primarily focuses on selecting four PEFT methods from PEFT library. Since all LLMs used in this work are casual language models. Following the guidance provided by PEFT library, there are three categories of PEFT methods to choose from: LoRA, prompt-based tuning, and $(IA)^3$. Finally, we select four PEFT methods, LoRA, p-tuning, prefix-tuning, and $(IA)^3$ to conduct the experiments. Figure \ref{fig:peft_arch} illustrates the working principles of these four PEFT methods on LLMs. Although \llama{} 2 uses Group Query Attention \cite{ainslie2023gqa}, to simplify the illustration, only Multi-head Attention is depicted in Figure \ref{fig:peft_arch}.

LoRA \cite{hu2021lora} focuses on $Query$ and $Key$ matrices within attention layers, as depicted in Figure \ref{fig:peft_arch}. LoRA incorporates additional trainable low-rank matrices to enable parameters updating. Specifically, for the initial weight matrix $W_0$, modification is performed through an additive operation $W_0 + BA$, where the rank of matrices $B$ and $A$ is considerably smaller than that of $W_0$, thereby reducing the number of trainable parameters.

Figure \ref{fig:peft_arch} demonstrates the architecture of two prompt-based tuning methods, p-tuning v2 \cite{liu2021p} and prefix-tuning \cite{li2021prefix}. Unlike static prompts commonly used in traditional prompt engineering, these prompts are dynamic and learnable, allowing for greater adaptability across various tasks. In p-tuning, trainable parameters are added to the input embeddings, while in prefix tuning, they are added to all layers, represented as $P_k$ in Figure \ref{fig:peft_arch}.

$(IA)^3$ \cite{liu2022few} stands for ``Infused Adapter by Inhibiting and Amplifying Inner Activation''. $(IA)^3$ introduces three learnable vectors: $l_k$, $l_v$, and $l_{ff}$, as shwon in Figure \ref{fig:peft_arch}. The vectors $l_k$ (for keys) and $l_v$ (for values) are used to rescale the attention keys and values, respectively, before applying the softmax. On the other hand, the vector $l_{ff}$ is employed to adjust the inner activations of the position-wise feed-forward networks. The attention mechanism with $(IA)^3$ is represented by the following formula:

$$\mathrm{softmax}\left(\frac{Q(l_\mathrm{k}\odot K^T)}{\sqrt{d_k}}\right)(l_\mathrm{v}\odot V)$$

Where $\odot$ denotes the element-wise multiplication,  $Q$ and $K$ are the $Query$ and $Key$ matrices, $V$ is the $Value$ matrix, and $d_k$ is the dimension of $Key$.

To ensure the fairness in the comparison of different PEFT methods on the same base model, we have maintained consistent magnitudes of trainable parameters among different PEFT methods by adjusting hyperparameters of different PEFT methods, shown in Table \ref{tab:appendix_peft}. Several challenges prevent prevent us from ensuring an exact match of trainable parameters across different PEFT methods in practice. Firstly, for $(IA)^3$, the number of trainable parameters is only determined by the structure of the base model, indicating $(IA)^3$ is impossible to have exactly the same trainable parameters as the other PEFT methods. Secondly, achieving a specific and exact number of trainable parameters is challenging. According to LoRA documentation, the rank should be set as an integer exponent of 2, which implies that adjusting the rank does not enable precise control over the desired number of trainable parameters. Overall, except for $(IA)^3$, we ensure that the number of trainable parameters in LoRA, p-tuning, and prefix-tuning remains within the same order of magnitude. Table \ref{tab:appendix_peft} also shows the setting of hyperparameters for different PEFT methods, taking \codellamasmall{} as an example. The setting of PEFT hyperparameters for other LLMs can be found in the source code \cite{repo}.

\begin{table}[h]
\caption{PEFT Hyperparameters setting of different PEFT Methods, taking \codellamasmall{} as an example.}
\label{tab:appendix_peft}
\resizebox{0.48\textwidth}{!}{%
\begin{tabular}{cccc}
\hline
PEFT Methods                   & Hyper-parameters      & value                 & Trainable parameters     \\ \hline
\multirow{2}{*}{$(IA)^3$}         & \multirow{2}{*}{None} & \multirow{2}{*}{None} & \multirow{2}{*}{$6.14 \times 10^5$} \\
                               &                       &                       &                          \\ \hline
\multirow{3}{*}{LoRA}          & rank                  & 32                    & \multirow{3}{*}{$1.68 \times 10^7$} \\
                               & lora alpha            & 16                    &                          \\
                               & lora dropout          & 0.05                  &                          \\ \hline
\multirow{3}{*}{p-tuning}      & virtual tokens        & 100                   & \multirow{3}{*}{$2.14 \times 10^7$} \\
                               & encoder hidden size   & 2048                  &                          \\
                               & encoder reparameterization type                  & MLP                  &                          \\ \hline
\multirow{2}{*}{prefix-tuning} & encoder hidden size   & 256                   & \multirow{2}{*}{$6.88 \times 10^7$} \\
                               & virtula tokens        & 100                   &                          \\ \hline
\end{tabular}
}
\end{table}

Indeed, there are several adapter-based PEFT methods available. However, the main focus of this work is to assess the effectiveness of PEFT methods, rather than comparing and determining the best method among all PEFT methods. Additionally, this work aims to simplify the application of PEFT methods by utilizing the Transformers and PEFT libraries for SFT, thereby enhancing the reusability of these methods for other software engineering tasks.

\subsection{Benchmarks and Evaluation Metrics of APR}

\subsubsection{Three existing benchmarks} 

Since \aprinst{} exclusively consists of Java instances, we specifically select APR benchmarks in Java for evaluation. In order to enable convenient comparison of fixing capabilities with previous work \cite{jiang2023impact}, this work specifically evaluates single-hunk bugs and utilizes the same three benchmarks as used in prior work: Defects4J \cite{just2014defects4j}, which comprises a collection of 217 single-hunk bugs carefully chosen by previous work \cite{jiang2023impact}, QuixBugs \cite{lin2017quixbugs}, which includes 40 bugs serves as a benchmark for well-known algorithm-related issues,  and HumanEval-Java \cite{jiang2023impact}, which covers 163 single-hunk Java bugs varying from simple errors to complex logical bugs. In all benchmarks, the location of the bug liens is known and label. Importantly, HumanEval-Java demonstrates a reduction of data leakage risks between pre-training data and benchmarks. All three benchmarks provide corresponding test suites for each bug, which are used to evaluate the correctness of the generated patches.

\subsubsection{Patch generation and validation}

In experiments, LLMs generate 10 candidate patches for each example in the benchmark. Recent studies \cite{noller2022trust} on developer preferences have shown that high-quality patches ranked within the top 10, generated within an acceptable time frame of 1 hour, are considered more valuable. Consequently, we individually validate the generated patches using the corresponding test suite to assess their correctness. A large number of experiments have been conducted in this work, generating over 7,000 plausible correct patches. Manually checking all of them would be excessively labor-intensive. Then, we manually checked a sample of the patches and all generated patch files have been published and are available for further examination.

Following other studies of execution-based evaluation, we used the $pass@k$ \cite{kulal2019spoc} to evaluate fixing capabilities. This metric generates $k$ code samples for each problem, and if any of the samples pass the unit tests, the problem is considered solved. In the subsequent experiments, we report the $pass@10$ metric for each benchmark.

\subsection{Implementation Details}

\begin{figure}[h]
  \vspace{-3mm}
  \centering
  \includegraphics[width=0.45\textwidth]{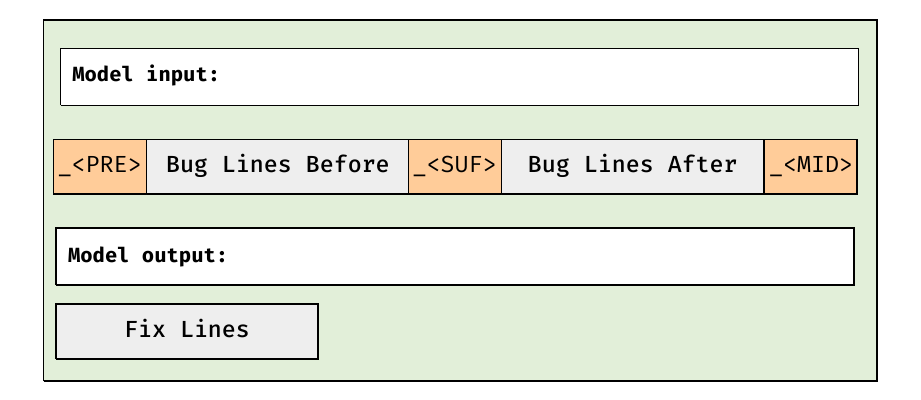}
  \caption{Infilling template of \codellamasmall{}}
  \label{fig:infill}
  \vspace{-4mm}
\end{figure}

This work also employs FMFT to explore the effectiveness and  efficiency of PEFT methods. Since it is not practical to directly obtain the fixing capability of no instruction-following LLMs, the benchmarks are transformed into an infilling task by employing the fill-in-the-middle task provided by \codellama to complete APR. In this infilling task, bug lines serve as lines that need to be completed by \codellama. As a result, we convert the buggy code from the three benchmarks into the infilling format depicted in Figure \ref{fig:infill} and feed it into \codellama for inference to generate patches for valuation. Additionally, during SFT, we record the usage of peak GPU memory and the number of trainable parameters in order to analyze the efficiency of PEFT in subsequent sections. Setting of trainning hyperparameters could be found in our source code \cite{repo}.

All pre-trained LLMs are downloaded from \textit{Hugging Face}. We conduct all experiments on a server equipped with Intel (R) Xeon (R) Platinum 8358 CPU, 1TB RAM, and 80G A800 GPU running Ubuntu 18.04. Additionally, we have released SFT and inference code to facilitate reproducibility by researchers \cite{repo}.

\section{Experiment Results}
\label{sec:experiment_results}

Firstly, we apply LoRA, p-tuning, prefix-tuning, $(IA)^3$, and FMFT on \codellamasmall{} to evaluate effectiveness of PEFT methods compared to no fine-tuning and other State-of-the-Art (SOTA) APR techniques. Secondly, we evaluate the impact of different PEFT methods and different LLMs on fixing capabilities. Lastly, we explore the influence of hyperparameters and the size of instruction dataset on the efficiency of the PEFT methods, aiming to address the following research question:

\begin{description}
    \item[\textbf{RQ1:}] How do PEFT methods effectively and efficiently improve the fixing capabilities of LLMs compared to no fine-tuning and SOTA APR techniques?
    \item[\textbf{RQ2:}] How do different PEFT methods improve the fixing capabilities of LLMs fine-tuned on different base LLMs?
    \item[\textbf{RQ3:}] How do PEFT hyperparameters and the size of instruction dataset affect fixing capabilities of LLMs on APR?
\end{description}

\subsection{RQ1: Fixing Capabilities and Computing Resource Consumption of PEFT}

Firstly, since our objective is to demonstrate the effectiveness of PEFT methods, we conduct a comparison between PEFT methods and the SOTA APR techniques provided in prior work \cite{jiang2023impact}. Secondly, we present PEFT methods improve fixing capabilities of \codellamasmall{} through fine-tuning, compared to the baseline, no fine-tuning. Lastly, as PEFT emphasizes the reduction of computational resource consumption, we conduct an analysis and comparison of computing resources consumption among different fine-tuning methods, including FMFT.

\subsubsection{Experiments Setup}

In experiments, we compare the computating resource consumption of different fine-tuning methods using specific hyperparameter settings to keep external environment settings consistent. We set \textit{epoch} to 5, \textit{maximum sequence length} to 1000, and \textit{learning  rate} to $1e^{-4}$, while further details can be found in our source code \cite{repo}. Following the guidance of PEFT methods, we assign \textit{rank} to 32 and \textit{lora  alpha} to 16 for LoRA, \textit{virtual  tokens} to 100, \textit{encoder hidden size} to 2048, and \textit{encoder reparameterization  type} to \textit{Multiple Layer  Perceptron (MLP)} for p-tuning, and \textit{virtual tokens} to 100 and \textit{encoder hidden size} to 256 for prefix-tuning. After SFT, we transform buggy code in benchmarks into instructions to guide LLMs in inferring and generating patches, which are validated by test cases, and $pass@k$ values are recorded, shown on Figure \ref{fig:arch} (f).

\subsubsection{Analysis of Fixing Capabilities}

\begin{table}[H]
  \caption{Number of fixed bugs of DL-based\cite{jiang2023impact}, LLM-based\cite{jiang2023impact} and PEFT methods on HumanEval-Java}
  \label{tab:rq1_sota}
  \resizebox{0.48\textwidth}{!}{%
  \begin{tabular}{ccccc}
  \hline
  \multirow{2}{*}{\begin{tabular}[c]{@{}c@{}}Base\\ Models\end{tabular}} & \multirow{2}{*}{\begin{tabular}[c]{@{}c@{}}Fine-tuning\\ Method\end{tabular}} & HumanEval-Java & HumanEval-Java & HumanEval-Java \\
  & & pass@1         & pass@5         & pass@10        \\ \hline
  CURE \cite{jiang2021cure} & FMFT & 7/163          & 15/163         & 18/163         \\
  RewardRepair \cite{ye2022neural} & FMFT  & 4/163          & 17/163         & 22/163         \\
  Recorder \cite{zhu2021syntax}& FMFT   & 5/163          & 7/163          & 11/163         \\
  INCODER-1B \cite{fried2022incoder}& FMFT & 38/163         & 54/163         & 64/163         \\
  INCODER-6B \cite{fried2022incoder}& FMFT  & 43/163         & 62/163         & 70/163         \\
  \codellamasmall  & $(IA)^3$ & 37/163         & 87/163         & 99/163         \\
  \codellamalarge  & $(IA)^3$ & 55/163         & 89/163         & 97/163        \\
  \deepseekfull & $(IA)^3$ & \cellcolor{morandiRed}{76/163}    & \cellcolor{morandiRed}{98/163}  & \cellcolor{morandiRed}{111/163}     \\ \hline
  \end{tabular}
  }
  \end{table}
  
We conduct a comparison of fixing capabilities between PEFT methods and the SOTA APR techniques based on previous work findings. It has been previously noted that Defects4J and QuixBugs present potential risks of data leakage \cite{jiang2023impact}. Therefore, we will compare the fixing capability on HumanEval-Java with previous APR techniques. To ensure consistency, we directly utilize the experimental results from previous research \cite{jiang2023impact} and present the comparison in Table \ref{tab:rq1_sota}. We select three DL-based methods, CURE \cite{jiang2021cure}, RewardRepair \cite{ye2022neural}, and Recorder \cite{zhu2021syntax}. Among these, RewardRepair demonstrates the highest fixing capability by successfully addressing 22 bugs out of 163 instances. Additionally, we evaluate the two most effective LLM-based APR technique, INCODER-1B \cite{fried2022incoder} and INCODER-6B \cite{fried2022incoder}, both fine-tuned with the APR dataset. Indeed, PEFT methods do not introduce any additional knowledge during fine-tuning compared to INCODER, as the dataset used for fine-tuning on INCODER-6B is the same one used to construct \aprinst.

According to the results in Table \ref{tab:rq1_sota}, INCODER-6B currently stands as the SOTA APR technique before this work, successfully fixing 70 bugs. LLMs with more parameters demonstrates enhanced fixing capability compared to DL-based methods. Additionally, Table \ref{tab:rq1_sota} displays fixing capabilities of the three LLMs employed in this work, fine-tuned with $(IA)^3$, on Humaneval-java. With $(IA)^3$, \codellamasmall{} achieves a higher fixing capability and resolves 99 bugs. The best model in this work, \deepseekfull{} with $(IA)^3$ fine-tuning, successfully fixes 111 bugs, which outperforms the previous SOTA, INCODER-6B, by fixing an additional 41 bugs, which is a 58\% improvement. Overall, PEFT methods indeeed leads to a improvement in fixing capability, enhancing it up to 58\% when compared to previous SOTA APR techniques.

\finding{
  As Table \ref{tab:rq1_sota} shown, the best model in this work, \deepseekfull{} with $(IA)^3$, fixes 58\% more bugs than the SOTA LLM-based APR technique, INCODER-6B with FMFT, on HumanEval-Java.   
}

\begin{table}[h]
\caption{Number of fixed bugs on different fine-tuning methods on \codellamasmall. (\%) represents the percentage of performance improvement compared to no fine-tuning(baseline).}
\label{tab:rq1_results}
\resizebox{0.46\textwidth}{!}{%
\begin{tabular}{ccccc}
\hline
\multirow{3}{*}{\begin{tabular}[c]{@{}c@{}}Fine-tuning\\ Method\end{tabular}} & \multicolumn{3}{c}{Benchmark} & \multirow{3}{*}{\begin{tabular}[c]{@{}c@{}}Total\\Fixed \\ Bugs\end{tabular}}\\ \cline{2-4}
& \begin{tabular}[c]{@{}c@{}}Humaneval-java\\ (pass@10)\end{tabular} & \begin{tabular}[c]{@{}c@{}}Defects4j v2.0\\ (pass@10)\end{tabular} & \begin{tabular}[c]{@{}c@{}}Quixbugs\\ (pass@10)\end{tabular} & \\ \hline
No Fine-tuning & 73/163  & 75/217    & 18/40 & 166/420(0\%)\\
FMFT           & 52/163  & 38/217    & 15/40 & 105/420(-36.7\%)\\ 
LoRA           & 98/163  & 75/217    & \cellcolor{morandiRed}{26/40} & 199/420(+19.9\%)\\
p-tuning       & 86/163  & 85/217    & 25/40 & 196/420(+18.1\%)\\
prefix-tuning  & 81/163  & 73/217    & 24/40 & 178/420(+7.2\%)\\
$(IA)^3$       & \cellcolor{morandiRed}{99/163} & \cellcolor{morandiRed}{95/217} & 24/40 & \cellcolor{morandiRed}{218/420(+31.3\%)}\\\hline
\end{tabular}}
\vspace{-3mm}
\end{table}

Table \ref{tab:rq1_results} is evident that PEFT methods have significantly improved the fixing capabilities compared to the baseline, no fine-tuning on \codellamasmall. $(IA)^3$ outperform the baseline by fixing 26 additional bugs on HumanEval-Java, 20 additional bugs on Defects4J, and 6 additional bugs on QuixBugs. This clearly demonstrates the effectiveness of PEFT methods in enhancing fixing capabilities on APR. However, it is important to note that FMFT performed inferior to both no fine-tuning and PEFT methods. This disparity can be attributed to the limited size of \aprinst, which contributes to overfitting issues with only 5 epochs. In an attempt to address this, we reduce the number of epochs to 3, but the improvement in fixing capabilities remained insufficient. Otherwise, during fine-tuning, we conduct $Adam$ optimizer to dynamically adjust the learning rate, but do not observe any benefits. Since FMFT requires over 300 times more trainable parameters than PEFT, the larger parameter size in FMFT may increase the risk of overfitting, which is consistent with the challenges observed in other studies when working with limited datasets. Overall, compared to no fine-tuning, the fixing capabilities of \codellamasmall{} with four PEFT methods have indeed improved, confirming the effectiveness of PEFT methods on APR.

\subsubsection{Analysis of Computing Resource Consumption}

\begin{figure}[h]
  \centering
  \includegraphics[width=0.48\textwidth]{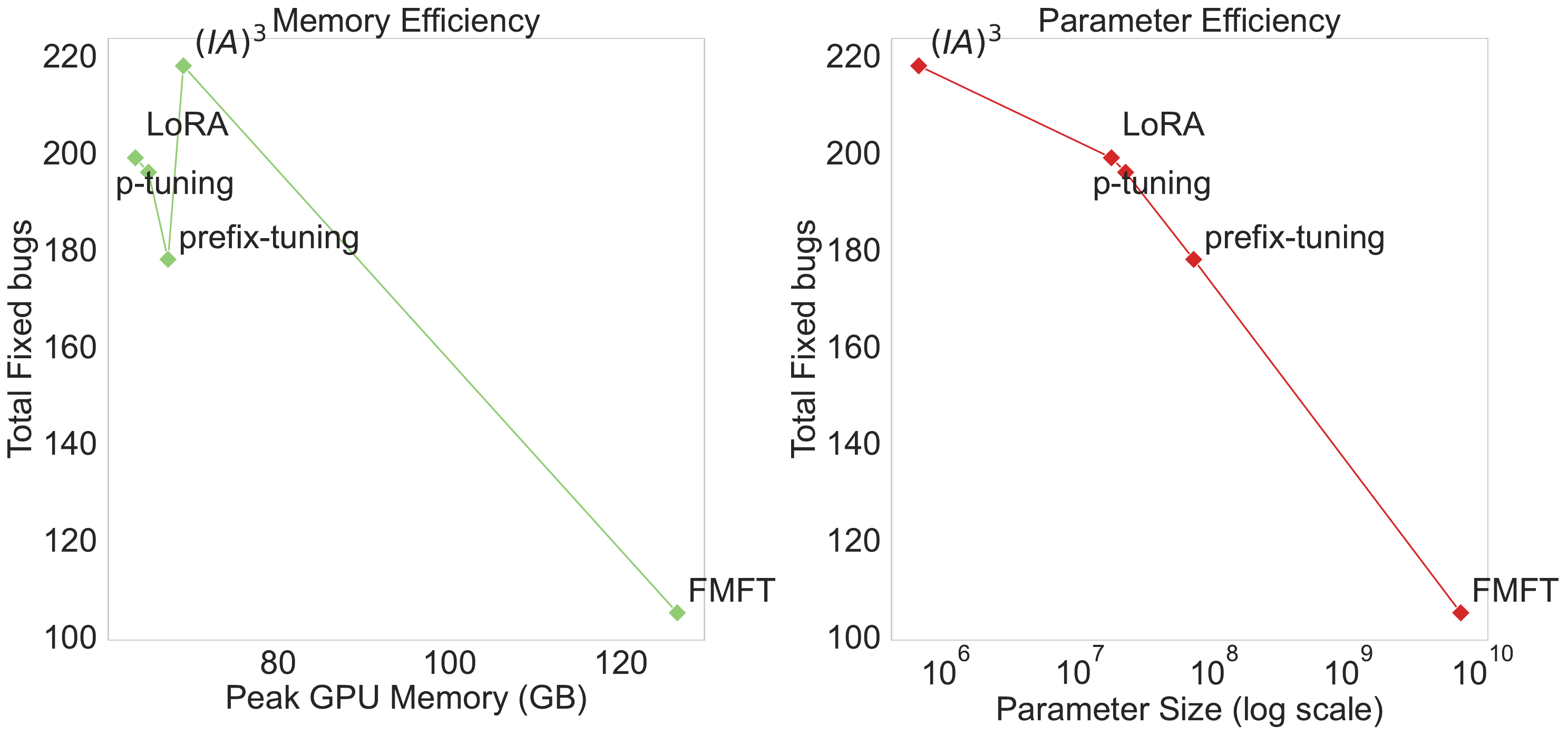}
  \caption{Efficiency of memory and parameters with different fine-tuning methods on \codellamasmall}
  \label{fig:rq1_memory_time}
  \vspace{-5mm}
\end{figure}

PEFT methods prioritize achieving performance improvement that is comparable to FMFT while aiming to reduce computational resource consumption, including peak GPU memory usage and the number of trainable parameters. Figure \ref{fig:rq1_memory_time} illustrates the relationship between computation resource consumption and fixing capabilities for different fine-tuning methods on \codellamasmall. FMFT consumes the largest number of trainable parameters and the highest peak GPU memory, 126.56GB where $(IA)^3$ consumes 68.91GB and LoRA consumes 63.30GB, 45.6\%-50.0\% less than FMFT, demonstrating the efficiency of PEFT methods in enhancing fixing capabilities on APR.
Intuitively, having fewer trainable parameters results in lower peak GPU memory consumption, as observed in the LoRA, p-tuning, and prefix-tuning methods, as Fiure \ref{fig:rq1_memory_time} shown. However, $(IA)^3$ has the fewest trainable parameters but consumes the most memory. This can be attributed to the principles of $(IA)^3$ discussed in the previous section \ref{sec:experiment_design}, where it introduces three additional scaling vectors, necessitating the use of three separate loss functions to update these vectors. This can lead to the need for storing more intermediate gradients, thus increasing memory usage. Nevertheless, $(IA)^3$ only requires approximately 8.86\% more memory compared to LoRA, while achieving 9.55\% more fixed bugs, showing the effectiveness and efficiency.

\finding{
  \codellamasmall{} fine-tuned with four PEFT methods has shown significantly improved fixing capabilities compared with no fine-tuning (fixing 7.2\%-31.3\% more fixed bugs) and reduces consumption of computational resources (45.6\%-50.0\% less peak GPU memory usage and around 97-10,00 times less trainable parameters) compared to FMFT.
}

\subsection{RQ2: Fix Capabilities in Different PEFT Methods Based on Different LLMs}

We conduct extensive experiments to investigate the factors that contribute to the improvement of fixing capabilities on \codellamasmall, \codellamalarge, \llamafull, and \deepseekfull, using four PEFT methods, including LoRA, p-tuning, prefix-tuning, and $(IA)^3$. Results are presented in Table \ref{tab:rq2_results}.

\begin{table}[h]
\caption{Number of fixed bugs on \codellamasmall, \codellamalarge, \llamafull, and \deepseekfull, fine-tuned with LoRA, p-tuning, prefix-tuning, and $(IA)^3$. \cellcolor{morandiRed}{Red} parts the highest fixing capability of different PEFT methods in the same base models, whicle \textbf{blod} indicates the highest fixing capability of the same PEFT method across distinct base models.}
\label{tab:rq2_results}
\resizebox{0.48\textwidth}{!}{%
\begin{tabular}{cccccc}
\hline
\multirow{3}{*}{Base Model}  & \multirow{3}{*}{PEFT Method} & \multicolumn{3}{c}{Benchmark} & \multirow{3}{*}{Total Fixed  Bugs} \\ \cline{3-5} 
& & \begin{tabular}[c]{@{}c@{}}HumanEval-Java\\ (pass@10)\end{tabular} & \begin{tabular}[c]{@{}c@{}}Defects4J v2.0\\ (pass@10)\end{tabular} & \begin{tabular}[c]{@{}c@{}}QuixBugs\\ (pass@10)\end{tabular} \\ \hline
\multirow{4}{*}{\codellamasmall} 
& LoRA           & 98/163   & 75/217   & \cellcolor{morandiRed}{26/40}  & 199/420 \\
& p-tuning       & 86/163   & 85/217   & 25/40  & 196/420 \\
& prefix-tuning  & 81/163   & 73/217   & 24/40  & 178/420 \\
& $(IA)^3$       & \cellcolor{morandiRed}{99/163} & \cellcolor{morandiRed}{95/217} & 24/40  & \cellcolor{morandiRed}{218/420} \\ \hline
\multirow{4}{*}{\codellamalarge}  
& LoRA           & 94/163   & 86/217   & 26/40  & 206/420 \\
& p-tuning       & 86/163   & 73/217   & 22/40  & 181/420 \\
& prefix-tuning  & 16/163   & 25/217   & 11/40  & 52/420  \\
& $(IA)^3$       & \cellcolor{morandiRed}{97/163} & \cellcolor{morandiRed}{97/217} & \cellcolor{morandiRed}{28/40} & \cellcolor{morandiRed}{222/420} \\ \hline
\multirow{4}{*}{\begin{tabular}[c]{@{}c@{}}\deepseekcoder \\ Base 6.7B\end{tabular}} 
& LoRA           & 109/163  & 92/217   & 30/40  & \textbf{231/420} \\
& p-tuning       & 108/163  & 84/217   & 33/40  & \textbf{225/420} \\
& prefix-tuning  & 98/163   & 73/217   & 30/40  & \textbf{201/420} \\
& $(IA)^3$       & \cellcolor{morandiRed}{111/163} & \cellcolor{morandiRed}{98/217} & \cellcolor{morandiRed}{34/40} & \cellcolor{morandiRed}{\textbf{243/420}} \\ \hline
\multirow{4}{*}{\llamafull{}}  
& LoRA           & \cellcolor{morandiRed}{24/163} & \cellcolor{morandiRed}{12/217} & \cellcolor{morandiRed}{2/40} & \cellcolor{morandiRed}{38/420} \\
& p-tuning       & 10/163   & 10/217   & 1/40   & 21/420 \\
& prefix-tuning  & 12/163   & 3/217    & 2/40   & 17/420 \\
& $(IA)^3$       & 17/163   & 3/217    & 2/40   & 22/420 \\ \hline
\end{tabular}}
\end{table}

\subsubsection{\codellamasmall{} vs \codellamalarge{}}

In terms of the number of total fixed bugs, both \codellamasmall{} and \codellamalarge{} demonstrate the highest fixing capability using $(IA)^3$. Notably, \codellamalarge{} fixes 4 more additional bugs than \codellamasmall{}, summarized in Table \ref{tab:rq2_results}. This finding aligns with the expectation that larger LLMs contribute to stronger fixing capabilities. However, we observe that, after applying p-tuning and prefix-tuning, the fixing capability of \codellamalarge{} is not as effective as that of \codellamasmall{} when using the same PEFT methods. Given the number of trainable parameters for p-tuning on \codellamalarge{} is 1.2 times that of \codellamasmall, while for prefix-tuning, it is 1.5 times. Hence, we consider that the decline in the fixing capability of p-tuning and prefix-tuning on \codellamalarge{} can be attributed to the increase in trainable parameters without a corresponding increase epoch, still 5 , which means LLMs may not be trained sufficiently. To further investigate this, additional experiments are conducted, and the results are presented in Table \ref{tab:rq2_13b_results}.

Table \ref{tab:rq2_13b_results} illustrates that increasing the epoch to 8 actually results in a decrease in the fixing capability for LoRA, while we will not continue to increase epoch. This indicates that 5 epochs may be sufficient, given that the parameter size of LoRA is smaller than that of p-tuning and prefix-tuning. In the case of p-tuning, when the epoch is increased to 10, \codellamalarge{} fixes 8 more bugs in HumanEval-Java, 11 more bugs in Defects4J, and 14 more bugs in QuixBugs, totaling 33 more bugs. On the other hand, prefix-tuning fixes 57 more bugs in HumanEval-Java, 26 more bugs in Defects4J, and 7 more bugs in QuixBugs, for a total of 100 additional bugs. However, the number of fixed bugs with prefix-tuning on \codellamalarge{} is still fewer than on \codellamasmall{}. It is possible that an epoch of 10 may still be insufficient, considering that prefix-tuning utilizes a greater number of trainable parameters compared to other PEFT methods. Due to limited computing resources, we will not delve further into this problem, although we have shown epoch affects the fixing capability of larger scale LLMs. Another possible factor is that the size of \aprinst{} is too small for such a large LLM, which implies that we need more data. Therefore, when employing a larger LLM for fine-tuning with p-tuning and prefix-tuning, it is advisable to appropriately increase the epoch to ensure adequate learning from the dataset.

\begin{table}[h]
\caption{Number of fixed bugs by increasing epoch on \codellamalarge}
\label{tab:rq2_13b_results}
\resizebox{0.48\textwidth}{!}{%
\begin{tabular}{ccccc}
\hline
\multirow{3}{*}{\begin{tabular}[c]{@{}c@{}}PEFT\\ Methods\end{tabular}} & \multirow{3}{*}{epoch} & \multicolumn{3}{c}{Benchmarks}\\ \cline{3-5} 
&  & \begin{tabular}[c]{@{}c@{}}HumanEval-Java\\ (pass@10)\end{tabular} & \begin{tabular}[c]{@{}c@{}}Defects4J v2.0\\ (pass@10)\end{tabular} & \begin{tabular}[c]{@{}c@{}}QuixBugs\\ (pass@10)\end{tabular} \\ \hline 
LoRA  & 5    & 94/163  & 86/217             & 26/40  \\
LoRA  & 8    & 89/163(-5)         & 82/217(-4)    & 27/40(+1)    \\ \hline
p-tuning  & 5   & 86/163  & 73/217             & 22/40  \\
p-tuning  & 10     & 94/163(+8)   & 84/217(+11)  & 36/40(+14)   \\\hline
prefix-tuning & 5    & 16/163                  & 25/217             & 11/40             \\ 
prefix-tuning & 10   & 73/163(+57)                  & 61/217(+26)             & 18/40(+7)             \\ \hline
\end{tabular}
}
\end{table}

\finding{
    \codellamalarge{} with LoRA, $(IA)^3$ and p-tuning fixes 7, 4, 18 more bugs than \codellamasmall{} with the same PEFT mthods. This demonstrates that increasing the scale of base models with PEFT can improve the fixing capability on APR. However, when using PEFT methods that require more parameters, such as p-tuning and prefix-tuning, we suggest appropriately increasing epoch ensures LLMs undergo sufficient training.
    }

\subsubsection{\codellamasmall{} vs \llamafull{}}

Table \ref{tab:rq2_results} clearly demonstrates that, even after fine-tuning, \llamafull{} continues to exhibit the lowest fixing capability. This outcome is unexpected, given that both \llamafull{} and \codellamasmall{} have the same parameter size. The primary contributing factor is the significant discrepancy between the text data used for pre-training \llamafull{} and the code data utilized for instruction-tuning. This finding emphasizes fine-tuning datasets of PEFT should be align with the pre-training data of LLMs in order to effectively enhance fixing capabilities on APR, as well as other code generation tasks.

\finding{\llamafull{} demonstrates the lowest fixing capability. We suggest to employ code language models  when utilizing PEFT for software engineering tasks.}

\subsubsection{\codellamasmall{} vs \deepseekfull{}}

According to Table \ref{tab:rq2_results}, \deepseekfull{} with $(IA)^3$ demonstrates superior fixing capability, outperforming \codellamasmall{} with $(IA)^3$ by fixing additional 25 bugs, which is also the highest fixing capability among all experiments conducted in this work. It is worth noting that both \codellamasmall{} and \deepseekfull{} have similar parameter sizes and the same model architecture. Hence, with the same parameter size, different LLMs exhibit varying degrees of improvement in the fixing capability through PEFT on APR. Furthermore, \deepseekfull{} fixes 21 more bugs than \codellamalarge{}, while the scale of \deepseekfull{} is smaller than that of \codellamalarge. In fact, four PEFT methods in this work achieved the highest fixing capability on \deepseekfull, as Table \ref{tab:rq2_results} \textbf{Bold Text} shows. This finding emphasizes the importance of selecting the appropriate base model when implementing the PEFT method. With the right choice, LLMs with a small scale can preform better than a larger-scale LLM.

\finding{\deepseekfull{} with $(IA)^3$ exhibits the highest fixing capability, successfully fixing 243 bugs totally, which demonstrates the fixing capability of PEFT methods relies on the selection of a suitable base model. We suggest utilizing \deepseekfull{} for APR and other SE tasks when implementing PEFT, may be a good starting point.}

\subsubsection{Impact of PEFT Methods}

\codellamasmall, \codellamalarge, and \deepseekfull{} all demonstrate the highest fixing capability with $(IA)^3$ from Table \ref{tab:rq2_results}. $(IA)^3$ also costs minimal parameter among four PEFT methods. Furthermore, $(IA)^3$ exhibits robustness as it does not require the configuration of any hyperparameters. Additionally, our experiments show that LoRA, closely following $(IA)^3$, not only enhances the fixing capability but also involves a relatively lower number of trainable parameters, which also contributes to the popularity of LoRA.

\begin{figure}[h]
  \centering
  \includegraphics[width=0.48\textwidth]{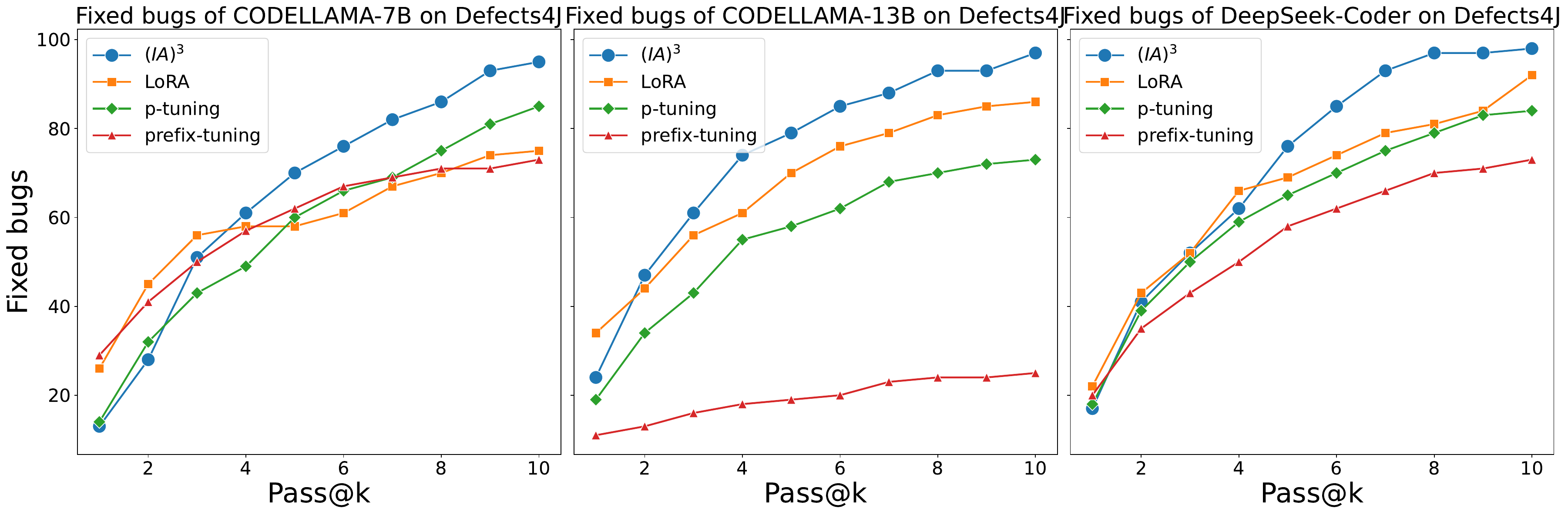}
  \caption{$pass@k$ on Defects4J with different PEFT methods}
  \label{fig:rq2_defects4j}
\end{figure}

Furthermore, we wonder to gain a deeper understanding of the significant enhancement of the fixing capability by $(IA)^3$. However, the original papers on PEFT methods do not explicitly provide proof of the performance improvement. Fortunately, this work focuses on a specific scenario, APR, which enables us to analyze the potential factors contributing to improving fixing capability by $(IA)^3$, at least for APR. By plotting the relationship between $pass@k$ and the number of fixed bugs for different PEFT methods, as shown in Figure \ref{fig:rq2_defects4j}, we have made interesting observations. Notably, $(IA)^3$ consistently exhibits lower fixing capabilities at $pass@1$ compared to other PEFT methods. For example, on Defects4J, we observe that $(IA)^3$ performs the lowest fixing capability at $pass@1$ when using \codellamasmall{}. However, starting at $pass@5$, $(IA)^3$ surpasses the other three PEFT methods and continues to improve until $pass@10$, ultimately achieving the best performance. This same trend is also observed on \codellamalarge{} and \deepseekfull{}.

\begin{figure}[h]
  \centering
    \vspace{-3mm}
  \includegraphics[width=0.48\textwidth]{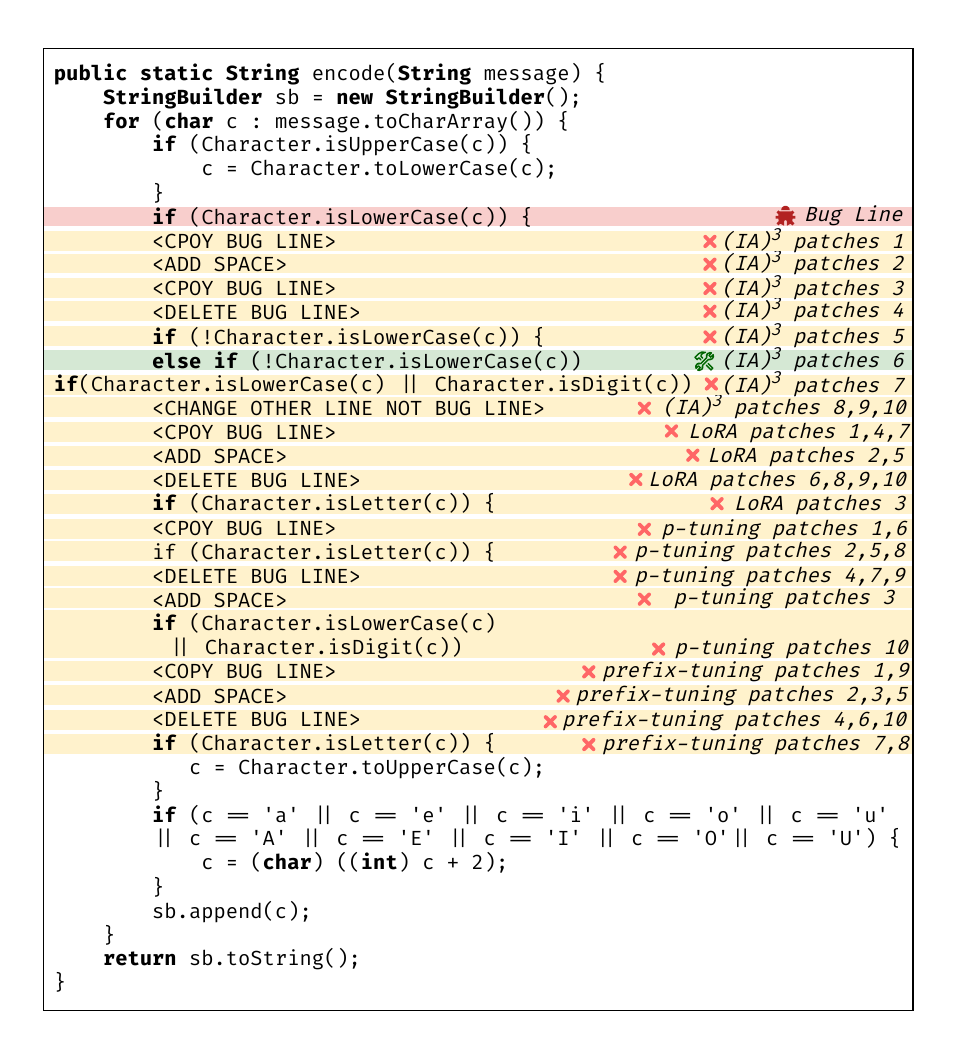}
  \caption{Instance of HumanEval-Java with different PEFT methods. This function represents an encoding method that involves swapping the case of all letters and replacing vowels in the message with the letter that appears two places ahead in the English alphabet.}
  \label{fig:instance}
\end{figure}

Figure \ref{fig:instance} presents an instance, \textit{ENCODE} of HumanEval-Java where only $(IA)^3$ successfully fixes the bug, while the other three PEFT methods failed to fix it on \deepseekfull. The bug line is highlighted in red in Figure \ref{fig:instance}. Figure \ref{fig:instance} illustrates ten candidate patches from each PEFT method, To aid understanding, some patches are represented as fixing behaviors such as <COPY BUG LINE>, <DELETE BUG LINE>, <ADD SPACE> and so on. Figure \ref{fig:instance} shows that $(IA)^3$ adopts a more conservative approach in the initial four patches, making minimal changes to the bug line, such as <COPY BUG LINE>. This also explains why $(IA)^3$ does not fix a significant number of bugs from pass@1 to pass@5. From the fifth patch, $(IA)^3$ employs a more extreme strategy by making substantial modifications to the bug line to explore the correct patch. The correct fix is accomplished in the sixth patch, and in the seventh patch, $(IA)^3$ continues to creatively modify the bug line. Starting from the eighth patch, $(IA)^3$ gradually extends modifications to lines outside of the bug line to complete the fix. 

For LoRA shown in Figure \ref{fig:instance}, most patches tend to make minimal changes to the bug line. It only modifies the bug line in the third patch and remains relatively conservative in the subsequent fixes. This aligns with the curve of Figure \ref{fig:rq2_defects4j}, demonstrating that the improvement of LoRA in the number of fixed bugs from pass@6 to pass@10 is not as significant as from pass@1 to pass@6. Given that this study primarily addresses single-hunk bugs, a more conservative fixing strategy may yield relatively fine results. P-tuning introduces significant modifications to the bug line in the  2nd, 3rd, 5th, and 8th patch, while adopting a conservative approach in the other patches. Although p-tuning enhances the creativity of LLMs, the improvement remains limited, resulting in the similarity of the 2nd, 5th, and 8th patch. Prefix-tuning only modifies the bug line in the 7th and 8th patch, but makes the same changes. The other patches generally leave the bug line unchanged. As a result, previous experiments indicate that prefix-tuning has the lowest fixing capability. From Figure \ref{fig:rq2_defects4j} and the instance in Figure \ref{fig:instance}, we hypothesize that $(IA)^3$ enhances the creativity of LLMs during fine-tuning which aims to generate more diverse outputs to explore more results as much as possible, thereby increasing the likelihood of generating the correct patch. However, it is essential to note that this conclusion is based on experimental observations and may not necessarily hold theoretical validity.

\begin{figure}[h]
  \centering
  \includegraphics[width=0.48\textwidth]{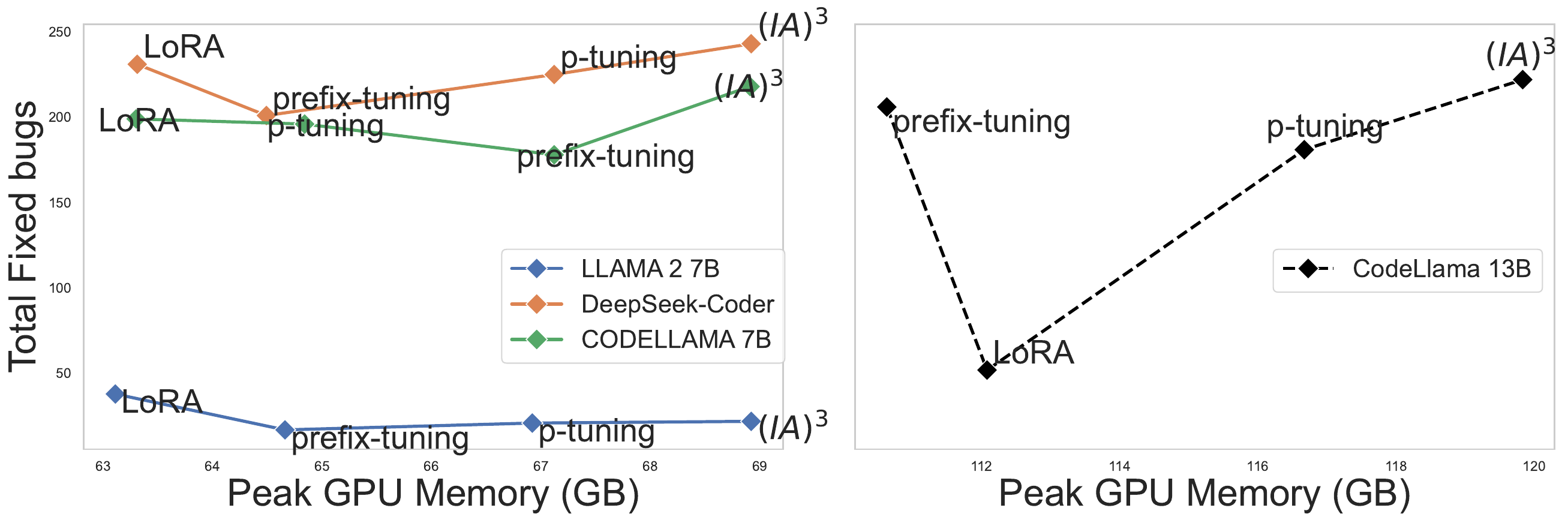}
  \caption{Peek GPU Memory on \codellamasmall, \codellamalarge, \llamafull{} and \deepseekfull{}}
  \label{fig:rq2_memory}
\end{figure}

In Figure \ref{fig:rq2_memory}, we present peak GPU memory usage of the four models during the training of the four PEFT methods. Among the applied fine-tuning methods, the peak GPU memory usage for the 7B model consistently remained below 70G, while the peak GPU memory consumption for fine-tuning the 13B model reached approximately 120G. Among the four evaluated PEFT methods, LoRA demonstrated the lowest memory consumption, followed by prefix-tuning and p-tuning, with $(IA)^3$ tending to exhibit the highest memory usage. Among 7B-scale LLMs, the difference in memory consumption between PEFT methods with the lowest and highest usage is about 4GB.

\finding{ 
    $(IA)^3$ achieves the highest fixing capability among all LLMs by improving the diversity of outputs, followed by LoRA, p-tuning, and prefix-tuning. We suggest leverage $(IA)^3$ and LoRA for fine-tuning as the first attempt on APR.
}

\subsection{RQ3: Exploration of Hyperparameters Setting and Sizes Setting of Instruction Dataset}

\subsubsection{Impact of Hyperparameters Setting}

When using PEFT methods, it is often necessary to configure specific hyperparameters to adjust the number of trainable parameters or the PEFT architecture. Examples of hyperparameters include \textit{rank} of LoRA, \textit{virtual tokens} of prefix-tuning, \textit{encoder hidden size} of p-tuning and so on. While $(IA)^3$ is the most effective PEFT method, but it lacks adjustable hyperparameters. However, LoRA, another popular PEFT method, exhibits the second fixing capability only to $(IA)^3$. Therefore, due to limited computational resources, we conduct experiments solely on \codellamasmall{} to explore \textit{rank} of LoRA.

\begin{table}[h]
\caption{Number of fixed bugs with different hyperparameters setting of LoRA \textit{rank} on \codellamasmall}
\label{tab:rq3_rank}
\resizebox{0.45\textwidth}{!}{%
\begin{tabular}{ccccc}
\hline
\multirow{3}{*}{\begin{tabular}[c]{@{}c@{}}LoRA\\ (rank)\end{tabular}} & \multicolumn{3}{c}{Benchmark} & \multirow{3}{*}{\begin{tabular}[c]{@{}c@{}}Total\\ Fixed \\ Bugs\end{tabular}} \\ \cline{2-4} 
& \begin{tabular}[c]{@{}c@{}}HumanEval-java\\ (pass@10)\end{tabular} & \begin{tabular}[c]{@{}c@{}}Defects4J v2.0\\ (pass@10)\end{tabular} & \begin{tabular}[c]{@{}c@{}}QuixBugs\\ (pass@10)\end{tabular} \\ \hline
8  & 90/163 & 74/217   & 25/40  & 189/420 \\
\cellcolor{morandiRed}{16}   & 99/163  &87/217    & 26/40  & \cellcolor{morandiRed}{212/420}   \\
32  & 98/163  & 75/217    & 26/40  &  199/420 \\
64  & 96/163  & 78/217    & 23/40  & 197/420 \\
128 & 98/163 & 78/217     & 26/40 & 202/420\\ \hline
\end{tabular}%
}
\end{table}

The guidelines of LoRA \cite{hu2021lora} suggest that the maximum $rank$ should not exceed $256$. As a result, we set the maximum $rank$ to $128$ and examined four different rank values: $8$, $16$, $32$, and $64$. We fix the other hyperparameters of LoRA, \textit{LoRA alpha} and \textit{LoRA dropout} at $16$ and $0.05$, respectively. The results are presented in Table \ref{tab:rq3_rank}. The $rank$ setting of $16$ demonstrates the highest fixing capability, fixing 13 more bugs compared to the experiment where the $rank$ is set to $32$, as conducted previously in this work. As shown in \ref{tab:rq3_rank}, the fixing capability initially increases and then decreases as the $rank$ increases. Thus, solely increasing the number of trainable parameters for PEFT does not necessarily enhance effectiveness.


\finding{Adjusting hyperparameters to increase the number of trainable parameters does not guarantee improvement. We suggest it maybe a good choice to set $rank$ to $16$ of LoRA as the initial configuration.}

\subsubsection{Impact of Instruction Dataset Size}

Using PEFT methods always entails limited computing resources. Consequently, we conduct an analysis to determine the impact of reducing the size of the training data on the fixing capabilities of PEFT methods, which is a way to reduce computing resource consumption. We select $(IA)^3$ and perform fine-tuning on \codellamasmall, which does not require the configuration of hyperparameters. \aprinst{} consists of 30K instructions. So, we conduct fine-tuning using 5K, 10K, 15K, 20K, 25K, and 30K instructions, respectively. The experimental results shown in Table \ref{tab:rq3_size} clearly indicate that the largest size, 30K, corresponds to the largest number of fixed bugs, validating the intuition that a larger dataset allows LLMs to acquire more knowledge and improve its fixing capability. This validation also suggests the effectiveness of \aprinst.

\begin{table}[h]
\caption{Number of fixed bugs with different sizes of \aprinst{} on \codellamasmall{} with $(IA)^3$. (\%) represents the percentage of performance degradation compared to the best result, highlighted by red.}
\label{tab:rq3_size}
\resizebox{0.48\textwidth}{!}{%
\begin{tabular}{ccccc}
\hline
\multirow{3}{*}{\begin{tabular}[c]{@{}c@{}}Size of\\ Dataset\end{tabular}} & \multicolumn{3}{c}{Benchmark} & \multirow{3}{*}{\begin{tabular}[c]{@{}c@{}}Total\\ Fixed \\ Bugs\end{tabular}} \\ \cline{2-4} 
& \begin{tabular}[c]{@{}c@{}}HumanEval-java\\ (pass@10)\end{tabular} & \begin{tabular}[c]{@{}c@{}}Defects4J v2.0\\ (pass@10)\end{tabular} & \begin{tabular}[c]{@{}c@{}}QuixBugs\\ (pass@10)\end{tabular} \\ \hline
5K   & 96/163 & 85/217 & 27/40  & 208/420 (-4.59\%) \\
10K  & 96/163 & 90/217 & 26/40  & 212/420 (-2.75\%) \\
15K  & 93/163 & 93/217 & 24/40  & 210/420 (-3.67\%) \\
20K  & 94/163 & 92/217 & 24/40  & 210/420 (-3.67\%) \\
25K  & 100/163 & 90/217 & 24/40 & 214/420 (-1.83\%) \\
30K  & 99/163 & 95/217 & 24/40  & \cellcolor{morandiRed}{218/420 (0\%)} \\ \hline
\end{tabular}
}
\vspace{-2mm}
\end{table}

Additionally, we found that when we reduced the dataset size to 25K, 16.7\% less than the full-size 30K, the fixing capability of LLMs only degraded by 1.83\%. Even when we further reduced the dataset size to 5K, 83.3\% less than the full-size 30K, the model fixes only 4.59\% fewer bugs. In situations where computational resources are limited, fine-tuning with $(IA)^3$ on a 5K dataset can still yield reasonably good results. This highlights the advantage of PEFT in resource-constrained scenarios, as it allows for minimizing training parameters and data while maintaining a fine performance.

\finding{The improvement in the fix capability of $(IA)^3$ does not significantly increase with a substantial expansion of the instruction dataset size. To balance computational resource consumption and the enhancement of fixing capability, we suggest researchers to start fine-tuning from a smaller-sized dataset and gradually increasing its size until the improvement in fix capability becomes insignificant.}

\section{DISCUSSION AND THREATS TO VALIDITY}
\label{sec:discussion}

This work mainly focuses the APR task in Java. Regrettably, no attempts have been made to extend these tasks to multiple programming languages, thereby limiting the applicability of the current methods to a broader range of software engineering challenges. Moreover, our concentration primarily lies on single-hunk APR tasks, neglecting to explore the fixing capabilities of PEFT in the multi-hunk APR task, which consequently leaves gaps in covering the entirety of APR scenarios. The hyperparameter settings employed for p-tuning and prefix-tuning in this work cannot guarantee optimality, suggesting a potential for improvement in certain experimental results. However, finding the optimal hyperparameters necessitates extensive experiments, rendering it challenging to determine the best configuration due to resource limitations. The patches passing test cases of benchmarks are ``plausible'' instead of ``correct'', caused by not ensuring complete line coverage of test cases, a common issue in software testing. Additionally, there exist quantization methods for PEFT, such as QLoRA \cite{dettmers2024qlora}, that can further reduce computational resource consumption. Nevertheless, these quantization methods often require additional environmental configurations, which have not been explored in this work. Nonetheless, quantization remains an exceptionally promising technique for further reducing resource consumption.

\section{RELATED WORK}
\label{sec:related_work}

\subsection{Parameter-Efficient Fine-Tuning for Software Engineering}  
\label{subsec:peft}

``Pre-training and fine-tuning'' \cite{zhao2023survey} enables LLMs to quickly adapt to SE tasks and improve the performance of LLMs. However, the parameter size of current LLMs is much larger, often reaching hundreds of billions, compared to previous models, which leads to significant computational resource consumption with FMFT. To reduce the resource consumption in fine-tuning, various techniques have been proposed, and one popular strategy is PEFT. Existing works have employed PEFT on SE tasks. The work \cite{zou2023comprehensive} proves the effectiveness PEFT methods by evaluating five PEFT methods on eight PTMs and four SE downstream tasks. Another work \cite{weyssow2023exploring} focuses on the effectiveness of PEFT methods on code generation, which illustrates the practicality of PEFT under a limited resources scenario, effectively mitigating the reliance on large and expensive computational infrastructures. The work \cite{wang2023one} explores parameter-efficient fine-tuning techniques for specializing LLMs for code search and code summarization, finding that parameter-efficient fine-tuning outperforms in-context learning. CodePrompt \cite{choi2023codeprompt} further propose a new PEFT method for SE tasks, to boosts code generation performance. The work \cite{shi2023towards} explores the impact of selectively freezing layers of the model. The evaluation of these studies primarily relies on similarity-based Exact Match and CodeBLEU metrics, which fail to capture the code actual executability.

\subsection{LLM-based Automated Program Repair}
\label{subsec:apr_on_llm}
With the advancements in pre-trained LLMs, recent research has begun exploring their usage on APR. An earlier study \cite{xia2023automated} comprehensively evaluates the performance of nine LLMs on APR tasks using various input forms. Compared to traditional APR techniques \cite{jobstmann2005program, khalilian2021cgenprog, long2015staged, wong2021varfix, chen2020contract, gao2021beyond, ghanbari2019practical}, LLMs have demonstrated distinct advantages, particularly in their ability to fix bugs without the need for manual provision of prior knowledge. Prior to the emergence of GPT, several works \cite{jin2023inferfix, paul2023enhancing, shirafuji2023program, wang2023rap} have proposed fine-tuning LLMs for APR, although the size of these LLMs does not exceed 7B. The study \cite{xia2023automated} further explains that employing models with larger parameters can enhance their repair ability. In a different research work \cite{jiang2023impact}, aside from directly employing LLMs for APR, ten Code Language Models (CLMs) were fine-tuned using APR datasets. The evaluation conducted on Defects4J, QuixBugs, and Humaneval-Java benchmarks demonstrates that CLMs, after fine-tuning with FMFT, exhibit improved fixing capabilities. These two studies emphasize the advantages of LLMs in APR tasks. However, LLMs used in these works were models released before 2022, with a maximum parameter size of 6B. The introduction of \textit{gpt3.5} in 2022 brought forth newly released LLMs that demonstrate even stronger performance. Several LLMs specifically designed for code-related tasks, such as StarCoder \cite{li2023starcoder}, \codellama{} \cite{codellama}, WizardCoder \cite{luo2024wizardcoder}, and DeepSeek-coder \cite{deepseek-coder}, have showcased impressive capabilities.

\section{CONCLUSION}
\label{sec:conclusion}

This work demonstrates the effectiveness and efficiency of PEFT in enhancing the fixing capability of LLMs on APR. Compared to the SOTA APR techniques, the best model presented in this work successfully fixes 58\% more bugs. Four LLMs and four PEFT methods are examined, resulting in sixteen sets of PEFT weights released soon. The impact of different base models and PEFT methods on the fixing capability is evaluated using three benchmarks. Specifically, an instance is provided to illustrate how $(IA)^3$ enhancing the creativity of LLM, outperformed the other three PEFT methods. Additionally, this work delves into the investigates of configuring hyperparameters of PEFT and reducing training data size to minimize computational resource consumption, while preserving the fixing capability of LLMs. Given the increasing scale of LLMs, we suggest researchers to consider employing PEFT for fine-tuning with limited computing resources. All datasets, code, and weights in this work will be available on \cite{repo}. Researchers can refer to the workflow outlined in this work and apply PEFT to other software engineering tasks, leveraging existing datasets, conveniently. In conclusion, we present a comprehensive roadmap for the application of Parameter-Effective Fine-Tuning on Automated Program Repair, highlighting its potential extension to other domains.

\begin{acks}
This research is supported by Open Project of Key Laboratory of Industrial Software Engineering and Application Technology, Ministry of Industry and Information Technology (HK202300064), the National Science Foundation of China under Grants U20A20173, 62125206 and 62402433, the National Key R\&D Program of China (2022YFF0902702) and the Key R\&D Program of Ningbo (2023Z235). This research is also supported in part by the Zhejiang Provincial Natural Science Foundation of China under Grant LQ24F020019. 

\end{acks}

\newpage
\bibliographystyle{ACM-Reference-Format}
\bibliography{cite}


\begin{thebibliography}{73}


\ifx \showCODEN    \undefined \def \showCODEN     #1{\unskip}     \fi
\ifx \showDOI      \undefined \def \showDOI       #1{#1}\fi
\ifx \showISBNx    \undefined \def \showISBNx     #1{\unskip}     \fi
\ifx \showISBNxiii \undefined \def \showISBNxiii  #1{\unskip}     \fi
\ifx \showISSN     \undefined \def \showISSN      #1{\unskip}     \fi
\ifx \showLCCN     \undefined \def \showLCCN      #1{\unskip}     \fi
\ifx \shownote     \undefined \def \shownote      #1{#1}          \fi
\ifx \showarticletitle \undefined \def \showarticletitle #1{#1}   \fi
\ifx \showURL      \undefined \def \showURL       {\relax}        \fi
\providecommand\bibfield[2]{#2}
\providecommand\bibinfo[2]{#2}
\providecommand\natexlab[1]{#1}
\providecommand\showeprint[2][]{arXiv:#2}

\bibitem[rep(2024)]%
        {repo}
 \bibinfo{year}{2024}\natexlab{}.
\newblock \bibinfo{howpublished}{\url{https://github.com/zjulgc/llmpeft4apr}}.
\newblock


\bibitem[hug(2024)]%
        {huggingface}
 \bibinfo{year}{2024}\natexlab{}.
\newblock \bibinfo{howpublished}{\url{https://huggingface.co}}.
\newblock


\bibitem[Ainslie et~al\mbox{.}(2023)]%
        {ainslie2023gqa}
\bibfield{author}{\bibinfo{person}{Joshua Ainslie}, \bibinfo{person}{James Lee-Thorp}, \bibinfo{person}{Michiel de Jong}, \bibinfo{person}{Yury Zemlyanskiy}, \bibinfo{person}{Federico Lebr{\'o}n}, {and} \bibinfo{person}{Sumit Sanghai}.} \bibinfo{year}{2023}\natexlab{}.
\newblock \showarticletitle{Gqa: Training generalized multi-query transformer models from multi-head checkpoints}.
\newblock \bibinfo{journal}{\emph{arXiv preprint arXiv:2305.13245}} (\bibinfo{year}{2023}).
\newblock


\bibitem[Benton et~al\mbox{.}(2020)]%
        {benton2020effectiveness}
\bibfield{author}{\bibinfo{person}{Samuel Benton}, \bibinfo{person}{Xia Li}, \bibinfo{person}{Yiling Lou}, {and} \bibinfo{person}{Lingming Zhang}.} \bibinfo{year}{2020}\natexlab{}.
\newblock \showarticletitle{On the effectiveness of unified debugging: An extensive study on 16 program repair systems}. In \bibinfo{booktitle}{\emph{Proceedings of the 35th IEEE/ACM International Conference on Automated Software Engineering}}. \bibinfo{pages}{907--918}.
\newblock


\bibitem[Chaudhary(2023)]%
        {codealpaca}
\bibfield{author}{\bibinfo{person}{Sahil Chaudhary}.} \bibinfo{year}{2023}\natexlab{}.
\newblock \bibinfo{title}{Code Alpaca: An Instruction-following LLaMA model for code generation}.
\newblock \bibinfo{howpublished}{\url{https://github.com/sahil280114/codealpaca}}.
\newblock


\bibitem[Chen et~al\mbox{.}(2020)]%
        {chen2020contract}
\bibfield{author}{\bibinfo{person}{Liushan Chen}, \bibinfo{person}{Yu Pei}, {and} \bibinfo{person}{Carlo~A Furia}.} \bibinfo{year}{2020}\natexlab{}.
\newblock \showarticletitle{Contract-based program repair without the contracts: An extended study}.
\newblock \bibinfo{journal}{\emph{IEEE Transactions on Software Engineering}} \bibinfo{volume}{47}, \bibinfo{number}{12} (\bibinfo{year}{2020}), \bibinfo{pages}{2841--2857}.
\newblock


\bibitem[Chen et~al\mbox{.}(2021)]%
        {chen2021evaluating}
\bibfield{author}{\bibinfo{person}{Mark Chen}, \bibinfo{person}{Jerry Tworek}, \bibinfo{person}{Heewoo Jun}, \bibinfo{person}{Qiming Yuan}, \bibinfo{person}{Henrique Ponde de~Oliveira Pinto}, \bibinfo{person}{Jared Kaplan}, \bibinfo{person}{Harri Edwards}, \bibinfo{person}{Yuri Burda}, \bibinfo{person}{Nicholas Joseph}, \bibinfo{person}{Greg Brockman}, {et~al\mbox{.}}} \bibinfo{year}{2021}\natexlab{}.
\newblock \showarticletitle{Evaluating large language models trained on code}.
\newblock \bibinfo{journal}{\emph{arXiv preprint arXiv:2107.03374}} (\bibinfo{year}{2021}).
\newblock


\bibitem[Chen et~al\mbox{.}(2019)]%
        {chen2019sequencer}
\bibfield{author}{\bibinfo{person}{Zimin Chen}, \bibinfo{person}{Steve Kommrusch}, \bibinfo{person}{Michele Tufano}, \bibinfo{person}{Louis-No{\"e}l Pouchet}, \bibinfo{person}{Denys Poshyvanyk}, {and} \bibinfo{person}{Martin Monperrus}.} \bibinfo{year}{2019}\natexlab{}.
\newblock \showarticletitle{Sequencer: Sequence-to-sequence learning for end-to-end program repair}.
\newblock \bibinfo{journal}{\emph{IEEE Transactions on Software Engineering}} \bibinfo{volume}{47}, \bibinfo{number}{9} (\bibinfo{year}{2019}), \bibinfo{pages}{1943--1959}.
\newblock


\bibitem[Choi and Lee(2023)]%
        {choi2023codeprompt}
\bibfield{author}{\bibinfo{person}{YunSeok Choi} {and} \bibinfo{person}{Jee-Hyong Lee}.} \bibinfo{year}{2023}\natexlab{}.
\newblock \showarticletitle{CodePrompt: Task-agnostic prefix tuning for program and language generation}. In \bibinfo{booktitle}{\emph{Findings of the Association for Computational Linguistics: ACL 2023}}. \bibinfo{pages}{5282--5297}.
\newblock


\bibitem[Dettmers et~al\mbox{.}(2024)]%
        {dettmers2024qlora}
\bibfield{author}{\bibinfo{person}{Tim Dettmers}, \bibinfo{person}{Artidoro Pagnoni}, \bibinfo{person}{Ari Holtzman}, {and} \bibinfo{person}{Luke Zettlemoyer}.} \bibinfo{year}{2024}\natexlab{}.
\newblock \showarticletitle{Qlora: Efficient finetuning of quantized llms}.
\newblock \bibinfo{journal}{\emph{Advances in Neural Information Processing Systems}}  \bibinfo{volume}{36} (\bibinfo{year}{2024}).
\newblock


\bibitem[Fried et~al\mbox{.}(2022)]%
        {fried2022incoder}
\bibfield{author}{\bibinfo{person}{Daniel Fried}, \bibinfo{person}{Armen Aghajanyan}, \bibinfo{person}{Jessy Lin}, \bibinfo{person}{Sida Wang}, \bibinfo{person}{Eric Wallace}, \bibinfo{person}{Freda Shi}, \bibinfo{person}{Ruiqi Zhong}, \bibinfo{person}{Wen-tau Yih}, \bibinfo{person}{Luke Zettlemoyer}, {and} \bibinfo{person}{Mike Lewis}.} \bibinfo{year}{2022}\natexlab{}.
\newblock \showarticletitle{Incoder: A generative model for code infilling and synthesis}.
\newblock \bibinfo{journal}{\emph{arXiv preprint arXiv:2204.05999}} (\bibinfo{year}{2022}).
\newblock


\bibitem[Gao et~al\mbox{.}(2021)]%
        {gao2021beyond}
\bibfield{author}{\bibinfo{person}{Xiang Gao}, \bibinfo{person}{Bo Wang}, \bibinfo{person}{Gregory~J Duck}, \bibinfo{person}{Ruyi Ji}, \bibinfo{person}{Yingfei Xiong}, {and} \bibinfo{person}{Abhik Roychoudhury}.} \bibinfo{year}{2021}\natexlab{}.
\newblock \showarticletitle{Beyond tests: Program vulnerability repair via crash constraint extraction}.
\newblock \bibinfo{journal}{\emph{ACM Transactions on Software Engineering and Methodology (TOSEM)}} \bibinfo{volume}{30}, \bibinfo{number}{2} (\bibinfo{year}{2021}), \bibinfo{pages}{1--27}.
\newblock


\bibitem[Ghanbari et~al\mbox{.}(2019)]%
        {ghanbari2019practical}
\bibfield{author}{\bibinfo{person}{Ali Ghanbari}, \bibinfo{person}{Samuel Benton}, {and} \bibinfo{person}{Lingming Zhang}.} \bibinfo{year}{2019}\natexlab{}.
\newblock \showarticletitle{Practical program repair via bytecode mutation}. In \bibinfo{booktitle}{\emph{Proceedings of the 28th ACM SIGSOFT International Symposium on Software Testing and Analysis}}. \bibinfo{pages}{19--30}.
\newblock


\bibitem[Guo et~al\mbox{.}(2024)]%
        {deepseek-coder}
\bibfield{author}{\bibinfo{person}{Daya Guo}, \bibinfo{person}{Qihao Zhu}, {and} \bibinfo{person}{Dejian Yang}.} \bibinfo{year}{2024}\natexlab{}.
\newblock \bibinfo{title}{DeepSeek-Coder: When the Large Language Model Meets Programming -- The Rise of Code Intelligence}.
\newblock
\newblock
\urldef\tempurl%
\url{https://arxiv.org/abs/2401.14196}
\showURL{%
\tempurl}


\bibitem[Honovich et~al\mbox{.}(2023)]%
        {honovich2023unnatural}
\bibfield{author}{\bibinfo{person}{Or Honovich}, \bibinfo{person}{Thomas Scialom}, \bibinfo{person}{Omer Levy}, {and} \bibinfo{person}{Timo Schick}.} \bibinfo{year}{2023}\natexlab{}.
\newblock \showarticletitle{Unnatural Instructions: Tuning Language Models with (Almost) No Human Labor}. In \bibinfo{booktitle}{\emph{The 61st Annual Meeting Of The Association For Computational Linguistics}}.
\newblock


\bibitem[Hu et~al\mbox{.}(2021)]%
        {hu2021lora}
\bibfield{author}{\bibinfo{person}{Edward~J Hu}, \bibinfo{person}{Yelong Shen}, \bibinfo{person}{Phillip Wallis}, \bibinfo{person}{Zeyuan Allen-Zhu}, \bibinfo{person}{Yuanzhi Li}, \bibinfo{person}{Shean Wang}, \bibinfo{person}{Lu Wang}, {and} \bibinfo{person}{Weizhu Chen}.} \bibinfo{year}{2021}\natexlab{}.
\newblock \showarticletitle{Lora: Low-rank adaptation of large language models}.
\newblock \bibinfo{journal}{\emph{arXiv preprint arXiv:2106.09685}} (\bibinfo{year}{2021}).
\newblock


\bibitem[Huang et~al\mbox{.}(2023a)]%
        {huang2023empirical}
\bibfield{author}{\bibinfo{person}{Kai Huang}, \bibinfo{person}{Xiangxin Meng}, \bibinfo{person}{Jian Zhang}, \bibinfo{person}{Yang Liu}, \bibinfo{person}{Wenjie Wang}, \bibinfo{person}{Shuhao Li}, {and} \bibinfo{person}{Yuqing Zhang}.} \bibinfo{year}{2023}\natexlab{a}.
\newblock \showarticletitle{An empirical study on fine-tuning large language models of code for automated program repair}. In \bibinfo{booktitle}{\emph{2023 38th IEEE/ACM International Conference on Automated Software Engineering (ASE)}}. IEEE, \bibinfo{pages}{1162--1174}.
\newblock


\bibitem[Huang et~al\mbox{.}(2023b)]%
        {huang2023survey}
\bibfield{author}{\bibinfo{person}{Kai Huang}, \bibinfo{person}{Zhengzi Xu}, \bibinfo{person}{Su Yang}, \bibinfo{person}{Hongyu Sun}, \bibinfo{person}{Xuejun Li}, \bibinfo{person}{Zheng Yan}, {and} \bibinfo{person}{Yuqing Zhang}.} \bibinfo{year}{2023}\natexlab{b}.
\newblock \showarticletitle{A Survey on Automated Program Repair Techniques}.
\newblock \bibinfo{journal}{\emph{arXiv preprint arXiv:2303.18184}} (\bibinfo{year}{2023}).
\newblock


\bibitem[Jiang et~al\mbox{.}(2018)]%
        {jiang2018shaping}
\bibfield{author}{\bibinfo{person}{Jiajun Jiang}, \bibinfo{person}{Yingfei Xiong}, \bibinfo{person}{Hongyu Zhang}, \bibinfo{person}{Qing Gao}, {and} \bibinfo{person}{Xiangqun Chen}.} \bibinfo{year}{2018}\natexlab{}.
\newblock \showarticletitle{Shaping program repair space with existing patches and similar code}. In \bibinfo{booktitle}{\emph{Proceedings of the 27th ACM SIGSOFT international symposium on software testing and analysis}}. \bibinfo{pages}{298--309}.
\newblock


\bibitem[Jiang et~al\mbox{.}(2023)]%
        {jiang2023impact}
\bibfield{author}{\bibinfo{person}{Nan Jiang}, \bibinfo{person}{Kevin Liu}, \bibinfo{person}{Thibaud Lutellier}, {and} \bibinfo{person}{Lin Tan}.} \bibinfo{year}{2023}\natexlab{}.
\newblock \showarticletitle{Impact of code language models on automated program repair}. In \bibinfo{booktitle}{\emph{2023 IEEE/ACM 45th International Conference on Software Engineering (ICSE)}}. IEEE, \bibinfo{pages}{1430--1442}.
\newblock


\bibitem[Jiang et~al\mbox{.}(2021)]%
        {jiang2021cure}
\bibfield{author}{\bibinfo{person}{Nan Jiang}, \bibinfo{person}{Thibaud Lutellier}, {and} \bibinfo{person}{Lin Tan}.} \bibinfo{year}{2021}\natexlab{}.
\newblock \showarticletitle{Cure: Code-aware neural machine translation for automatic program repair}. In \bibinfo{booktitle}{\emph{2021 IEEE/ACM 43rd International Conference on Software Engineering (ICSE)}}. IEEE, \bibinfo{pages}{1161--1173}.
\newblock


\bibitem[Jin et~al\mbox{.}(2023)]%
        {jin2023inferfix}
\bibfield{author}{\bibinfo{person}{Matthew Jin}, \bibinfo{person}{Syed Shahriar}, \bibinfo{person}{Michele Tufano}, \bibinfo{person}{Xin Shi}, \bibinfo{person}{Shuai Lu}, \bibinfo{person}{Neel Sundaresan}, {and} \bibinfo{person}{Alexey Svyatkovskiy}.} \bibinfo{year}{2023}\natexlab{}.
\newblock \showarticletitle{Inferfix: End-to-end program repair with llms}. In \bibinfo{booktitle}{\emph{Proceedings of the 31st ACM Joint European Software Engineering Conference and Symposium on the Foundations of Software Engineering}}. \bibinfo{pages}{1646--1656}.
\newblock


\bibitem[Jobstmann et~al\mbox{.}(2005)]%
        {jobstmann2005program}
\bibfield{author}{\bibinfo{person}{Barbara Jobstmann}, \bibinfo{person}{Andreas Griesmayer}, {and} \bibinfo{person}{Roderick Bloem}.} \bibinfo{year}{2005}\natexlab{}.
\newblock \showarticletitle{Program repair as a game}. In \bibinfo{booktitle}{\emph{Computer Aided Verification: 17th International Conference, CAV 2005, Edinburgh, Scotland, UK, July 6-10, 2005. Proceedings 17}}. Springer, \bibinfo{pages}{226--238}.
\newblock


\bibitem[Just et~al\mbox{.}(2014)]%
        {just2014defects4j}
\bibfield{author}{\bibinfo{person}{Ren{\'e} Just}, \bibinfo{person}{Darioush Jalali}, {and} \bibinfo{person}{Michael~D Ernst}.} \bibinfo{year}{2014}\natexlab{}.
\newblock \showarticletitle{Defects4J: A database of existing faults to enable controlled testing studies for Java programs}. In \bibinfo{booktitle}{\emph{Proceedings of the 2014 international symposium on software testing and analysis}}. \bibinfo{pages}{437--440}.
\newblock


\bibitem[Khalilian et~al\mbox{.}(2021)]%
        {khalilian2021cgenprog}
\bibfield{author}{\bibinfo{person}{Alireza Khalilian}, \bibinfo{person}{Ahmad Baraani-Dastjerdi}, {and} \bibinfo{person}{Bahman Zamani}.} \bibinfo{year}{2021}\natexlab{}.
\newblock \showarticletitle{CGenProg: Adaptation of cartesian genetic programming with migration and opposite guesses for automatic repair of software regression faults}.
\newblock \bibinfo{journal}{\emph{Expert Systems with Applications}}  \bibinfo{volume}{169} (\bibinfo{year}{2021}), \bibinfo{pages}{114503}.
\newblock


\bibitem[Kulal et~al\mbox{.}(2019)]%
        {kulal2019spoc}
\bibfield{author}{\bibinfo{person}{Sumith Kulal}, \bibinfo{person}{Panupong Pasupat}, \bibinfo{person}{Kartik Chandra}, \bibinfo{person}{Mina Lee}, \bibinfo{person}{Oded Padon}, \bibinfo{person}{Alex Aiken}, {and} \bibinfo{person}{Percy~S Liang}.} \bibinfo{year}{2019}\natexlab{}.
\newblock \showarticletitle{Spoc: Search-based pseudocode to code}.
\newblock \bibinfo{journal}{\emph{Advances in Neural Information Processing Systems}}  \bibinfo{volume}{32} (\bibinfo{year}{2019}).
\newblock


\bibitem[Le et~al\mbox{.}(2016)]%
        {le2016history}
\bibfield{author}{\bibinfo{person}{Xuan Bach~D Le}, \bibinfo{person}{David Lo}, {and} \bibinfo{person}{Claire Le~Goues}.} \bibinfo{year}{2016}\natexlab{}.
\newblock \showarticletitle{History driven program repair}. In \bibinfo{booktitle}{\emph{2016 IEEE 23rd international conference on software analysis, evolution, and reengineering (SANER)}}, Vol.~\bibinfo{volume}{1}. IEEE, \bibinfo{pages}{213--224}.
\newblock


\bibitem[Le~Goues et~al\mbox{.}(2011)]%
        {le2011genprog}
\bibfield{author}{\bibinfo{person}{Claire Le~Goues}, \bibinfo{person}{ThanhVu Nguyen}, \bibinfo{person}{Stephanie Forrest}, {and} \bibinfo{person}{Westley Weimer}.} \bibinfo{year}{2011}\natexlab{}.
\newblock \showarticletitle{Genprog: A generic method for automatic software repair}.
\newblock \bibinfo{journal}{\emph{Ieee transactions on software engineering}} \bibinfo{volume}{38}, \bibinfo{number}{1} (\bibinfo{year}{2011}), \bibinfo{pages}{54--72}.
\newblock


\bibitem[Li et~al\mbox{.}(2023)]%
        {li2023starcoder}
\bibfield{author}{\bibinfo{person}{Raymond Li}, \bibinfo{person}{Loubna~Ben Allal}, \bibinfo{person}{Yangtian Zi}, \bibinfo{person}{Niklas Muennighoff}, \bibinfo{person}{Denis Kocetkov}, \bibinfo{person}{Chenghao Mou}, \bibinfo{person}{Marc Marone}, \bibinfo{person}{Christopher Akiki}, \bibinfo{person}{Jia Li}, \bibinfo{person}{Jenny Chim}, {et~al\mbox{.}}} \bibinfo{year}{2023}\natexlab{}.
\newblock \showarticletitle{Starcoder: may the source be with you!}
\newblock \bibinfo{journal}{\emph{arXiv preprint arXiv:2305.06161}} (\bibinfo{year}{2023}).
\newblock


\bibitem[Li and Liang(2021)]%
        {li2021prefix}
\bibfield{author}{\bibinfo{person}{Xiang~Lisa Li} {and} \bibinfo{person}{Percy Liang}.} \bibinfo{year}{2021}\natexlab{}.
\newblock \showarticletitle{Prefix-tuning: Optimizing continuous prompts for generation}.
\newblock \bibinfo{journal}{\emph{arXiv preprint arXiv:2101.00190}} (\bibinfo{year}{2021}).
\newblock


\bibitem[Lin et~al\mbox{.}(2017)]%
        {lin2017quixbugs}
\bibfield{author}{\bibinfo{person}{Derrick Lin}, \bibinfo{person}{James Koppel}, \bibinfo{person}{Angela Chen}, {and} \bibinfo{person}{Armando Solar-Lezama}.} \bibinfo{year}{2017}\natexlab{}.
\newblock \showarticletitle{QuixBugs: A multi-lingual program repair benchmark set based on the Quixey Challenge}. In \bibinfo{booktitle}{\emph{Proceedings Companion of the 2017 ACM SIGPLAN international conference on systems, programming, languages, and applications: software for humanity}}. \bibinfo{pages}{55--56}.
\newblock


\bibitem[Liu et~al\mbox{.}(2022)]%
        {liu2022few}
\bibfield{author}{\bibinfo{person}{Haokun Liu}, \bibinfo{person}{Derek Tam}, \bibinfo{person}{Mohammed Muqeeth}, \bibinfo{person}{Jay Mohta}, \bibinfo{person}{Tenghao Huang}, \bibinfo{person}{Mohit Bansal}, {and} \bibinfo{person}{Colin~A Raffel}.} \bibinfo{year}{2022}\natexlab{}.
\newblock \showarticletitle{Few-shot parameter-efficient fine-tuning is better and cheaper than in-context learning}.
\newblock \bibinfo{journal}{\emph{Advances in Neural Information Processing Systems}}  \bibinfo{volume}{35} (\bibinfo{year}{2022}), \bibinfo{pages}{1950--1965}.
\newblock


\bibitem[Liu et~al\mbox{.}(2024b)]%
        {liu2024your}
\bibfield{author}{\bibinfo{person}{Jiawei Liu}, \bibinfo{person}{Chunqiu~Steven Xia}, \bibinfo{person}{Yuyao Wang}, {and} \bibinfo{person}{Lingming Zhang}.} \bibinfo{year}{2024}\natexlab{b}.
\newblock \showarticletitle{Is your code generated by chatgpt really correct? rigorous evaluation of large language models for code generation}.
\newblock \bibinfo{journal}{\emph{Advances in Neural Information Processing Systems}}  \bibinfo{volume}{36} (\bibinfo{year}{2024}).
\newblock


\bibitem[Liu et~al\mbox{.}(2019)]%
        {liu2019you}
\bibfield{author}{\bibinfo{person}{Kui Liu}, \bibinfo{person}{Anil Koyuncu}, \bibinfo{person}{Tegawend{\'e}~F Bissyand{\'e}}, \bibinfo{person}{Dongsun Kim}, \bibinfo{person}{Jacques Klein}, {and} \bibinfo{person}{Yves Le~Traon}.} \bibinfo{year}{2019}\natexlab{}.
\newblock \showarticletitle{You cannot fix what you cannot find! an investigation of fault localization bias in benchmarking automated program repair systems}. In \bibinfo{booktitle}{\emph{2019 12th IEEE conference on software testing, validation and verification (ICST)}}. IEEE, \bibinfo{pages}{102--113}.
\newblock


\bibitem[Liu et~al\mbox{.}(2020)]%
        {liu2020efficiency}
\bibfield{author}{\bibinfo{person}{Kui Liu}, \bibinfo{person}{Shangwen Wang}, \bibinfo{person}{Anil Koyuncu}, \bibinfo{person}{Kisub Kim}, \bibinfo{person}{Tegawend{\'e}~F Bissyand{\'e}}, \bibinfo{person}{Dongsun Kim}, \bibinfo{person}{Peng Wu}, \bibinfo{person}{Jacques Klein}, \bibinfo{person}{Xiaoguang Mao}, {and} \bibinfo{person}{Yves~Le Traon}.} \bibinfo{year}{2020}\natexlab{}.
\newblock \showarticletitle{On the efficiency of test suite based program repair: A systematic assessment of 16 automated repair systems for java programs}. In \bibinfo{booktitle}{\emph{Proceedings of the ACM/IEEE 42nd International Conference on Software Engineering}}. \bibinfo{pages}{615--627}.
\newblock


\bibitem[Liu et~al\mbox{.}(2024a)]%
        {liu2024delving}
\bibfield{author}{\bibinfo{person}{Shuo Liu}, \bibinfo{person}{Jacky Keung}, \bibinfo{person}{Zhen Yang}, \bibinfo{person}{Fang Liu}, \bibinfo{person}{Qilin Zhou}, {and} \bibinfo{person}{Yihan Liao}.} \bibinfo{year}{2024}\natexlab{a}.
\newblock \showarticletitle{Delving into Parameter-Efficient Fine-Tuning in Code Change Learning: An Empirical Study}.
\newblock \bibinfo{journal}{\emph{arXiv preprint arXiv:2402.06247}} (\bibinfo{year}{2024}).
\newblock


\bibitem[Liu et~al\mbox{.}(2021)]%
        {liu2021p}
\bibfield{author}{\bibinfo{person}{Xiao Liu}, \bibinfo{person}{Kaixuan Ji}, \bibinfo{person}{Yicheng Fu}, \bibinfo{person}{Weng~Lam Tam}, \bibinfo{person}{Zhengxiao Du}, \bibinfo{person}{Zhilin Yang}, {and} \bibinfo{person}{Jie Tang}.} \bibinfo{year}{2021}\natexlab{}.
\newblock \showarticletitle{P-tuning v2: Prompt tuning can be comparable to fine-tuning universally across scales and tasks}.
\newblock \bibinfo{journal}{\emph{arXiv preprint arXiv:2110.07602}} (\bibinfo{year}{2021}).
\newblock


\bibitem[Long and Rinard(2015)]%
        {long2015staged}
\bibfield{author}{\bibinfo{person}{Fan Long} {and} \bibinfo{person}{Martin Rinard}.} \bibinfo{year}{2015}\natexlab{}.
\newblock \showarticletitle{Staged program repair with condition synthesis}. In \bibinfo{booktitle}{\emph{Proceedings of the 2015 10th Joint Meeting on Foundations of Software Engineering}}. \bibinfo{pages}{166--178}.
\newblock


\bibitem[Lu et~al\mbox{.}(2021)]%
        {lu2021codexglue}
\bibfield{author}{\bibinfo{person}{Shuai Lu}, \bibinfo{person}{Daya Guo}, \bibinfo{person}{Shuo Ren}, \bibinfo{person}{Junjie Huang}, \bibinfo{person}{Alexey Svyatkovskiy}, \bibinfo{person}{Ambrosio Blanco}, \bibinfo{person}{Colin Clement}, \bibinfo{person}{Dawn Drain}, \bibinfo{person}{Daxin Jiang}, \bibinfo{person}{Duyu Tang}, {et~al\mbox{.}}} \bibinfo{year}{2021}\natexlab{}.
\newblock \showarticletitle{Codexglue: A machine learning benchmark dataset for code understanding and generation}.
\newblock \bibinfo{journal}{\emph{arXiv preprint arXiv:2102.04664}} (\bibinfo{year}{2021}).
\newblock


\bibitem[Luo et~al\mbox{.}(2024)]%
        {luo2024wizardcoder}
\bibfield{author}{\bibinfo{person}{Ziyang Luo}, \bibinfo{person}{Can Xu}, \bibinfo{person}{Pu Zhao}, \bibinfo{person}{Qingfeng Sun}, \bibinfo{person}{Xiubo Geng}, \bibinfo{person}{Wenxiang Hu}, \bibinfo{person}{Chongyang Tao}, \bibinfo{person}{Jing Ma}, \bibinfo{person}{Qingwei Lin}, {and} \bibinfo{person}{Daxin Jiang}.} \bibinfo{year}{2024}\natexlab{}.
\newblock \showarticletitle{WizardCoder: Empowering Code Large Language Models with Evol-Instruct}. In \bibinfo{booktitle}{\emph{The Twelfth International Conference on Learning Representations}}.
\newblock
\urldef\tempurl%
\url{https://openreview.net/forum?id=UnUwSIgK5W}
\showURL{%
\tempurl}


\bibitem[Lutellier et~al\mbox{.}(2020)]%
        {lutellier2020coconut}
\bibfield{author}{\bibinfo{person}{Thibaud Lutellier}, \bibinfo{person}{Hung~Viet Pham}, \bibinfo{person}{Lawrence Pang}, \bibinfo{person}{Yitong Li}, \bibinfo{person}{Moshi Wei}, {and} \bibinfo{person}{Lin Tan}.} \bibinfo{year}{2020}\natexlab{}.
\newblock \showarticletitle{Coconut: combining context-aware neural translation models using ensemble for program repair}. In \bibinfo{booktitle}{\emph{Proceedings of the 29th ACM SIGSOFT international symposium on software testing and analysis}}. \bibinfo{pages}{101--114}.
\newblock


\bibitem[Maharana et~al\mbox{.}(2024)]%
        {maharana2024mathbb}
\bibfield{author}{\bibinfo{person}{Adyasha Maharana}, \bibinfo{person}{Prateek Yadav}, {and} \bibinfo{person}{Mohit Bansal}.} \bibinfo{year}{2024}\natexlab{}.
\newblock \showarticletitle{$\mathbb{D}^2$ Pruning: Message Passing for Balancing Diversity \& Difficulty in Data Pruning}. In \bibinfo{booktitle}{\emph{The Twelfth International Conference on Learning Representations}}.
\newblock


\bibitem[Mangrulkar et~al\mbox{.}(2022)]%
        {peft}
\bibfield{author}{\bibinfo{person}{Sourab Mangrulkar}, \bibinfo{person}{Sylvain Gugger}, \bibinfo{person}{Lysandre Debut}, \bibinfo{person}{Younes Belkada}, \bibinfo{person}{Sayak Paul}, {and} \bibinfo{person}{Benjamin Bossan}.} \bibinfo{year}{2022}\natexlab{}.
\newblock \bibinfo{title}{PEFT: State-of-the-art Parameter-Efficient Fine-Tuning methods}.
\newblock \bibinfo{howpublished}{\url{https://github.com/huggingface/peft}}.
\newblock


\bibitem[Nguyen et~al\mbox{.}(2013)]%
        {nguyen2013semfix}
\bibfield{author}{\bibinfo{person}{Hoang Duong~Thien Nguyen}, \bibinfo{person}{Dawei Qi}, \bibinfo{person}{Abhik Roychoudhury}, {and} \bibinfo{person}{Satish Chandra}.} \bibinfo{year}{2013}\natexlab{}.
\newblock \showarticletitle{Semfix: Program repair via semantic analysis}. In \bibinfo{booktitle}{\emph{2013 35th International Conference on Software Engineering (ICSE)}}. IEEE, \bibinfo{pages}{772--781}.
\newblock


\bibitem[Noller et~al\mbox{.}(2022)]%
        {noller2022trust}
\bibfield{author}{\bibinfo{person}{Yannic Noller}, \bibinfo{person}{Ridwan Shariffdeen}, \bibinfo{person}{Xiang Gao}, {and} \bibinfo{person}{Abhik Roychoudhury}.} \bibinfo{year}{2022}\natexlab{}.
\newblock \showarticletitle{Trust enhancement issues in program repair}. In \bibinfo{booktitle}{\emph{Proceedings of the 44th International Conference on Software Engineering}}. \bibinfo{pages}{2228--2240}.
\newblock


\bibitem[OpenAI(2023)]%
        {chatgpt}
\bibfield{author}{\bibinfo{person}{OpenAI}.} \bibinfo{year}{2023}\natexlab{}.
\newblock \bibinfo{howpublished}{\url{https://openai.com/blog/chatgpt}}.
\newblock


\bibitem[Papineni et~al\mbox{.}(2002)]%
        {papineni2002bleu}
\bibfield{author}{\bibinfo{person}{Kishore Papineni}, \bibinfo{person}{Salim Roukos}, \bibinfo{person}{Todd Ward}, {and} \bibinfo{person}{Wei-Jing Zhu}.} \bibinfo{year}{2002}\natexlab{}.
\newblock \showarticletitle{Bleu: a method for automatic evaluation of machine translation}. In \bibinfo{booktitle}{\emph{Proceedings of the 40th annual meeting of the Association for Computational Linguistics}}. \bibinfo{pages}{311--318}.
\newblock


\bibitem[Paul et~al\mbox{.}(2023)]%
        {paul2023enhancing}
\bibfield{author}{\bibinfo{person}{Rishov Paul}, \bibinfo{person}{Md~Mohib Hossain}, \bibinfo{person}{Mohammed~Latif Siddiq}, \bibinfo{person}{Masum Hasan}, \bibinfo{person}{Anindya Iqbal}, {and} \bibinfo{person}{Joanna~CS Santos}.} \bibinfo{year}{2023}\natexlab{}.
\newblock \showarticletitle{Enhancing Automated Program Repair through Fine-tuning and Prompt Engineering}.
\newblock \bibinfo{journal}{\emph{arXiv preprint arXiv:2304.07840}} (\bibinfo{year}{2023}).
\newblock


\bibitem[Radford et~al\mbox{.}(2019)]%
        {radford2019language}
\bibfield{author}{\bibinfo{person}{Alec Radford}, \bibinfo{person}{Jeffrey Wu}, \bibinfo{person}{Rewon Child}, \bibinfo{person}{David Luan}, \bibinfo{person}{Dario Amodei}, \bibinfo{person}{Ilya Sutskever}, {et~al\mbox{.}}} \bibinfo{year}{2019}\natexlab{}.
\newblock \showarticletitle{Language models are unsupervised multitask learners}.
\newblock \bibinfo{journal}{\emph{OpenAI blog}} \bibinfo{volume}{1}, \bibinfo{number}{8} (\bibinfo{year}{2019}), \bibinfo{pages}{9}.
\newblock


\bibitem[Rozi{\`{e}}re et~al\mbox{.}(2023)]%
        {codellama}
\bibfield{author}{\bibinfo{person}{Baptiste Rozi{\`{e}}re}, \bibinfo{person}{Jonas Gehring}, {and} \bibinfo{person}{Fabian Gloeckle}.} \bibinfo{year}{2023}\natexlab{}.
\newblock \showarticletitle{Code Llama: Open Foundation Models for Code}.
\newblock \bibinfo{journal}{\emph{CoRR}}  \bibinfo{volume}{abs/2308.12950} (\bibinfo{year}{2023}).
\newblock
\urldef\tempurl%
\url{https://doi.org/10.48550/ARXIV.2308.12950}
\showDOI{\tempurl}
\showeprint[arXiv]{2308.12950}


\bibitem[Shi et~al\mbox{.}(2023)]%
        {shi2023towards}
\bibfield{author}{\bibinfo{person}{Ensheng Shi}, \bibinfo{person}{Yanlin Wang}, \bibinfo{person}{Hongyu Zhang}, \bibinfo{person}{Lun Du}, \bibinfo{person}{Shi Han}, \bibinfo{person}{Dongmei Zhang}, {and} \bibinfo{person}{Hongbin Sun}.} \bibinfo{year}{2023}\natexlab{}.
\newblock \showarticletitle{Towards efficient fine-tuning of pre-trained code models: An experimental study and beyond}. In \bibinfo{booktitle}{\emph{Proceedings of the 32nd ACM SIGSOFT International Symposium on Software Testing and Analysis}}. \bibinfo{pages}{39--51}.
\newblock


\bibitem[Shirafuji et~al\mbox{.}(2023)]%
        {shirafuji2023program}
\bibfield{author}{\bibinfo{person}{Atsushi Shirafuji}, \bibinfo{person}{Md~Mostafizer Rahman}, \bibinfo{person}{Md~Faizul~Ibne Amin}, {and} \bibinfo{person}{Yutaka Watanobe}.} \bibinfo{year}{2023}\natexlab{}.
\newblock \showarticletitle{Program repair with minimal edits using codet5}. In \bibinfo{booktitle}{\emph{2023 12th International Conference on Awareness Science and Technology (iCAST)}}. IEEE, \bibinfo{pages}{178--184}.
\newblock


\bibitem[Touvron et~al\mbox{.}(2023a)]%
        {touvron2023llama}
\bibfield{author}{\bibinfo{person}{Hugo Touvron}, \bibinfo{person}{Thibaut Lavril}, \bibinfo{person}{Gautier Izacard}, \bibinfo{person}{Xavier Martinet}, \bibinfo{person}{Marie-Anne Lachaux}, \bibinfo{person}{Timoth{\'e}e Lacroix}, \bibinfo{person}{Baptiste Rozi{\`e}re}, \bibinfo{person}{Naman Goyal}, \bibinfo{person}{Eric Hambro}, \bibinfo{person}{Faisal Azhar}, {et~al\mbox{.}}} \bibinfo{year}{2023}\natexlab{a}.
\newblock \showarticletitle{Llama: Open and efficient foundation language models}.
\newblock \bibinfo{journal}{\emph{arXiv preprint arXiv:2302.13971}} (\bibinfo{year}{2023}).
\newblock


\bibitem[Touvron et~al\mbox{.}(2023b)]%
        {llama2}
\bibfield{author}{\bibinfo{person}{Hugo Touvron}, \bibinfo{person}{Louis Martin}, {and} \bibinfo{person}{Kevin Stone}.} \bibinfo{year}{2023}\natexlab{b}.
\newblock \showarticletitle{Llama 2: Open Foundation and Fine-Tuned Chat Models}.
\newblock \bibinfo{journal}{\emph{CoRR}}  \bibinfo{volume}{abs/2307.09288} (\bibinfo{year}{2023}).
\newblock
\urldef\tempurl%
\url{https://doi.org/10.48550/ARXIV.2307.09288}
\showDOI{\tempurl}
\showeprint[arXiv]{2307.09288}


\bibitem[Tufano et~al\mbox{.}(2019)]%
        {tufano2019empirical}
\bibfield{author}{\bibinfo{person}{Michele Tufano}, \bibinfo{person}{Cody Watson}, \bibinfo{person}{Gabriele Bavota}, \bibinfo{person}{Massimiliano~Di Penta}, \bibinfo{person}{Martin White}, {and} \bibinfo{person}{Denys Poshyvanyk}.} \bibinfo{year}{2019}\natexlab{}.
\newblock \showarticletitle{An empirical study on learning bug-fixing patches in the wild via neural machine translation}.
\newblock \bibinfo{journal}{\emph{ACM Transactions on Software Engineering and Methodology (TOSEM)}} \bibinfo{volume}{28}, \bibinfo{number}{4} (\bibinfo{year}{2019}), \bibinfo{pages}{1--29}.
\newblock


\bibitem[Wang et~al\mbox{.}(2023a)]%
        {wang2023one}
\bibfield{author}{\bibinfo{person}{Deze Wang}, \bibinfo{person}{Boxing Chen}, \bibinfo{person}{Shanshan Li}, \bibinfo{person}{Wei Luo}, \bibinfo{person}{Shaoliang Peng}, \bibinfo{person}{Wei Dong}, {and} \bibinfo{person}{Xiangke Liao}.} \bibinfo{year}{2023}\natexlab{a}.
\newblock \showarticletitle{One adapter for all programming languages? adapter tuning for code search and summarization}. In \bibinfo{booktitle}{\emph{2023 IEEE/ACM 45th International Conference on Software Engineering (ICSE)}}. IEEE, \bibinfo{pages}{5--16}.
\newblock


\bibitem[Wang et~al\mbox{.}(2020)]%
        {wang2020automated}
\bibfield{author}{\bibinfo{person}{Shangwen Wang}, \bibinfo{person}{Ming Wen}, \bibinfo{person}{Bo Lin}, \bibinfo{person}{Hongjun Wu}, \bibinfo{person}{Yihao Qin}, \bibinfo{person}{Deqing Zou}, \bibinfo{person}{Xiaoguang Mao}, {and} \bibinfo{person}{Hai Jin}.} \bibinfo{year}{2020}\natexlab{}.
\newblock \showarticletitle{Automated patch correctness assessment: How far are we?}. In \bibinfo{booktitle}{\emph{Proceedings of the 35th IEEE/ACM International Conference on Automated Software Engineering}}. \bibinfo{pages}{968--980}.
\newblock


\bibitem[Wang et~al\mbox{.}(2023c)]%
        {wang2023rap}
\bibfield{author}{\bibinfo{person}{Weishi Wang}, \bibinfo{person}{Yue Wang}, \bibinfo{person}{Shafiq Joty}, {and} \bibinfo{person}{Steven~CH Hoi}.} \bibinfo{year}{2023}\natexlab{c}.
\newblock \showarticletitle{Rap-gen: Retrieval-augmented patch generation with codet5 for automatic program repair}. In \bibinfo{booktitle}{\emph{Proceedings of the 31st ACM Joint European Software Engineering Conference and Symposium on the Foundations of Software Engineering}}. \bibinfo{pages}{146--158}.
\newblock


\bibitem[Wang et~al\mbox{.}(2023b)]%
        {wang2023self}
\bibfield{author}{\bibinfo{person}{Yizhong Wang}, \bibinfo{person}{Yeganeh Kordi}, \bibinfo{person}{Swaroop Mishra}, \bibinfo{person}{Alisa Liu}, \bibinfo{person}{Noah~A Smith}, \bibinfo{person}{Daniel Khashabi}, {and} \bibinfo{person}{Hannaneh Hajishirzi}.} \bibinfo{year}{2023}\natexlab{b}.
\newblock \showarticletitle{Self-Instruct: Aligning Language Models with Self-Generated Instructions}. In \bibinfo{booktitle}{\emph{ACL}}.
\newblock


\bibitem[Wei et~al\mbox{.}(2021)]%
        {wei2021finetuned}
\bibfield{author}{\bibinfo{person}{Jason Wei}, \bibinfo{person}{Maarten Bosma}, \bibinfo{person}{Vincent~Y Zhao}, \bibinfo{person}{Kelvin Guu}, \bibinfo{person}{Adams~Wei Yu}, \bibinfo{person}{Brian Lester}, \bibinfo{person}{Nan Du}, \bibinfo{person}{Andrew~M Dai}, {and} \bibinfo{person}{Quoc~V Le}.} \bibinfo{year}{2021}\natexlab{}.
\newblock \showarticletitle{Finetuned language models are zero-shot learners}.
\newblock \bibinfo{journal}{\emph{arXiv preprint arXiv:2109.01652}} (\bibinfo{year}{2021}).
\newblock


\bibitem[Wei et~al\mbox{.}(2024)]%
        {wei2023magicoder}
\bibfield{author}{\bibinfo{person}{Yuxiang Wei}, \bibinfo{person}{Zhe Wang}, \bibinfo{person}{Jiawei Liu}, \bibinfo{person}{Yifeng Ding}, {and} \bibinfo{person}{Lingming Zhang}.} \bibinfo{year}{2024}\natexlab{}.
\newblock \showarticletitle{Magicoder: Empowering code generation with oss-instruct}.
\newblock  (\bibinfo{year}{2024}).
\newblock


\bibitem[Weyssow et~al\mbox{.}(2023)]%
        {weyssow2023exploring}
\bibfield{author}{\bibinfo{person}{Martin Weyssow}, \bibinfo{person}{Xin Zhou}, \bibinfo{person}{Kisub Kim}, \bibinfo{person}{David Lo}, {and} \bibinfo{person}{Houari Sahraoui}.} \bibinfo{year}{2023}\natexlab{}.
\newblock \showarticletitle{Exploring parameter-efficient fine-tuning techniques for code generation with large language models}.
\newblock \bibinfo{journal}{\emph{arXiv preprint arXiv:2308.10462}} (\bibinfo{year}{2023}).
\newblock


\bibitem[Wolf et~al\mbox{.}(2020)]%
        {wolf-etal-2020-transformers}
\bibfield{author}{\bibinfo{person}{Thomas Wolf}, \bibinfo{person}{Lysandre Debut}, \bibinfo{person}{Victor Sanh}, \bibinfo{person}{Julien Chaumond}, \bibinfo{person}{Clement Delangue}, \bibinfo{person}{Anthony Moi}, \bibinfo{person}{Pierric Cistac}, \bibinfo{person}{Tim Rault}, \bibinfo{person}{Rémi Louf}, \bibinfo{person}{Morgan Funtowicz}, \bibinfo{person}{Joe Davison}, \bibinfo{person}{Sam Shleifer}, \bibinfo{person}{Patrick von Platen}, \bibinfo{person}{Clara Ma}, \bibinfo{person}{Yacine Jernite}, \bibinfo{person}{Julien Plu}, \bibinfo{person}{Canwen Xu}, \bibinfo{person}{Teven~Le Scao}, \bibinfo{person}{Sylvain Gugger}, \bibinfo{person}{Mariama Drame}, \bibinfo{person}{Quentin Lhoest}, {and} \bibinfo{person}{Alexander~M. Rush}.} \bibinfo{year}{2020}\natexlab{}.
\newblock \showarticletitle{Transformers: State-of-the-Art Natural Language Processing}. In \bibinfo{booktitle}{\emph{Proceedings of the 2020 Conference on Empirical Methods in Natural Language Processing: System Demonstrations}}. \bibinfo{publisher}{Association for Computational Linguistics}, \bibinfo{address}{Online}, \bibinfo{pages}{38--45}.
\newblock
\urldef\tempurl%
\url{https://www.aclweb.org/anthology/2020.emnlp-demos.6}
\showURL{%
\tempurl}


\bibitem[Wong et~al\mbox{.}(2021)]%
        {wong2021varfix}
\bibfield{author}{\bibinfo{person}{Chu-Pan Wong}, \bibinfo{person}{Priscila Santiesteban}, \bibinfo{person}{Christian K{\"a}stner}, {and} \bibinfo{person}{Claire Le~Goues}.} \bibinfo{year}{2021}\natexlab{}.
\newblock \showarticletitle{VarFix: balancing edit expressiveness and search effectiveness in automated program repair}. In \bibinfo{booktitle}{\emph{Proceedings of the 29th ACM joint meeting on European software engineering conference and symposium on the foundations of software engineering}}. \bibinfo{pages}{354--366}.
\newblock


\bibitem[Xia et~al\mbox{.}(2023)]%
        {xia2023automated}
\bibfield{author}{\bibinfo{person}{Chunqiu~Steven Xia}, \bibinfo{person}{Yuxiang Wei}, {and} \bibinfo{person}{Lingming Zhang}.} \bibinfo{year}{2023}\natexlab{}.
\newblock \showarticletitle{Automated program repair in the era of large pre-trained language models}. In \bibinfo{booktitle}{\emph{2023 IEEE/ACM 45th International Conference on Software Engineering (ICSE)}}. IEEE, \bibinfo{pages}{1482--1494}.
\newblock


\bibitem[Xia and Zhang(2022)]%
        {xia2022less}
\bibfield{author}{\bibinfo{person}{Chunqiu~Steven Xia} {and} \bibinfo{person}{Lingming Zhang}.} \bibinfo{year}{2022}\natexlab{}.
\newblock \showarticletitle{Less training, more repairing please: revisiting automated program repair via zero-shot learning}. In \bibinfo{booktitle}{\emph{Proceedings of the 30th ACM Joint European Software Engineering Conference and Symposium on the Foundations of Software Engineering}}. \bibinfo{pages}{959--971}.
\newblock


\bibitem[Xuan et~al\mbox{.}(2016)]%
        {xuan2016nopol}
\bibfield{author}{\bibinfo{person}{Jifeng Xuan}, \bibinfo{person}{Matias Martinez}, \bibinfo{person}{Favio Demarco}, \bibinfo{person}{Maxime Clement}, \bibinfo{person}{Sebastian~Lamelas Marcote}, \bibinfo{person}{Thomas Durieux}, \bibinfo{person}{Daniel Le~Berre}, {and} \bibinfo{person}{Martin Monperrus}.} \bibinfo{year}{2016}\natexlab{}.
\newblock \showarticletitle{Nopol: Automatic repair of conditional statement bugs in java programs}.
\newblock \bibinfo{journal}{\emph{IEEE Transactions on Software Engineering}} \bibinfo{volume}{43}, \bibinfo{number}{1} (\bibinfo{year}{2016}), \bibinfo{pages}{34--55}.
\newblock


\bibitem[Ye et~al\mbox{.}(2022)]%
        {ye2022neural}
\bibfield{author}{\bibinfo{person}{He Ye}, \bibinfo{person}{Matias Martinez}, {and} \bibinfo{person}{Martin Monperrus}.} \bibinfo{year}{2022}\natexlab{}.
\newblock \showarticletitle{Neural program repair with execution-based backpropagation}. In \bibinfo{booktitle}{\emph{Proceedings of the 44th international conference on software engineering}}. \bibinfo{pages}{1506--1518}.
\newblock


\bibitem[Zhang et~al\mbox{.}(2023)]%
        {zhang2023survey}
\bibfield{author}{\bibinfo{person}{Quanjun Zhang}, \bibinfo{person}{Chunrong Fang}, \bibinfo{person}{Yuxiang Ma}, \bibinfo{person}{Weisong Sun}, {and} \bibinfo{person}{Zhenyu Chen}.} \bibinfo{year}{2023}\natexlab{}.
\newblock \showarticletitle{A survey of learning-based automated program repair}.
\newblock \bibinfo{journal}{\emph{ACM Transactions on Software Engineering and Methodology}} \bibinfo{volume}{33}, \bibinfo{number}{2} (\bibinfo{year}{2023}), \bibinfo{pages}{1--69}.
\newblock


\bibitem[Zhang et~al\mbox{.}(2024)]%
        {zhang2024systematic}
\bibfield{author}{\bibinfo{person}{Quanjun Zhang}, \bibinfo{person}{Chunrong Fang}, \bibinfo{person}{Yang Xie}, \bibinfo{person}{YuXiang Ma}, \bibinfo{person}{Weisong Sun}, {and} \bibinfo{person}{Yun Yang~Zhenyu Chen}.} \bibinfo{year}{2024}\natexlab{}.
\newblock \showarticletitle{A Systematic Literature Review on Large Language Models for Automated Program Repair}.
\newblock \bibinfo{journal}{\emph{arXiv preprint arXiv:2405.01466}} (\bibinfo{year}{2024}).
\newblock


\bibitem[Zhao et~al\mbox{.}(2023)]%
        {zhao2023survey}
\bibfield{author}{\bibinfo{person}{Wayne~Xin Zhao}, \bibinfo{person}{Kun Zhou}, \bibinfo{person}{Junyi Li}, \bibinfo{person}{Tianyi Tang}, \bibinfo{person}{Xiaolei Wang}, \bibinfo{person}{Yupeng Hou}, \bibinfo{person}{Yingqian Min}, \bibinfo{person}{Beichen Zhang}, \bibinfo{person}{Junjie Zhang}, \bibinfo{person}{Zican Dong}, {et~al\mbox{.}}} \bibinfo{year}{2023}\natexlab{}.
\newblock \showarticletitle{A survey of large language models}.
\newblock \bibinfo{journal}{\emph{arXiv preprint arXiv:2303.18223}} (\bibinfo{year}{2023}).
\newblock


\bibitem[Zhu et~al\mbox{.}(2021)]%
        {zhu2021syntax}
\bibfield{author}{\bibinfo{person}{Qihao Zhu}, \bibinfo{person}{Zeyu Sun}, \bibinfo{person}{Yuan-an Xiao}, \bibinfo{person}{Wenjie Zhang}, \bibinfo{person}{Kang Yuan}, \bibinfo{person}{Yingfei Xiong}, {and} \bibinfo{person}{Lu Zhang}.} \bibinfo{year}{2021}\natexlab{}.
\newblock \showarticletitle{A syntax-guided edit decoder for neural program repair}. In \bibinfo{booktitle}{\emph{Proceedings of the 29th ACM joint meeting on European software engineering conference and symposium on the foundations of software engineering}}. \bibinfo{pages}{341--353}.
\newblock


\bibitem[Zou et~al\mbox{.}(2023)]%
        {zou2023comprehensive}
\bibfield{author}{\bibinfo{person}{Wentao Zou}, \bibinfo{person}{Qi Li}, \bibinfo{person}{Jidong Ge}, \bibinfo{person}{Chuanyi Li}, \bibinfo{person}{Xiaoyu Shen}, \bibinfo{person}{Liguo Huang}, {and} \bibinfo{person}{Bin Luo}.} \bibinfo{year}{2023}\natexlab{}.
\newblock \showarticletitle{A Comprehensive Evaluation of Parameter-Efficient Fine-Tuning on Software Engineering Tasks}.
\newblock \bibinfo{journal}{\emph{arXiv preprint arXiv:2312.15614}} (\bibinfo{year}{2023}).
\newblock


\end{thebibliography}
\end{document}